\documentclass[useAMS,usenatbib]{mn2e}

\usepackage{verbatim} 
\usepackage{natbib} 
\usepackage{amsmath} 
\usepackage{amsbsy}
\usepackage{amssymb}
\usepackage{mathrsfs} 
\usepackage{lscape} 
\usepackage{graphicx}
\usepackage{epstopdf}
\usepackage{deluxetable}

\newcommand{\Lya}{${\rm Ly}\alpha$}

\title[Baryonic Halo Assembly]{The Baryonic Assembly of Dark Matter Halos}
\author[Faucher-Gigu\`ere, Kere\v{s}, \& Ma]{Claude-Andr\'e Faucher-Gigu\`ere\thanks{Miller Fellow;
cgiguere@berkeley.edu}, Du\v{s}an Kere\v{s}\thanks{Hubble Fellow}, Chung-Pei Ma\\
Department of Astronomy and Theoretical Astrophysics Center, University of California, Berkeley, CA 94720-3411, USA.}

\begin{document}
\maketitle


\begin{abstract}
We use a suite of cosmological hydrodynamic simulations to systematically quantify the accretion rates of baryons into dark matter halos and the resulting baryon mass fractions, as a function of halo mass, redshift, and baryon type (including cold and hot gas). 
We find that the net baryonic accretion rates through the virial radius are sensitive to galactic outflows and explore a range of outflow parameters to illustrate the effects. 
We show that the cold gas accretion rate is in general not a simple universal factor of the dark matter accretion rate, and that galactic winds can cause star formation rates to deviate significantly from the external gas accretion rates, both via gas ejection and re-accretion. 
Furthermore, galactic winds can inject enough energy and momentum in the surrounding medium to slow down accretion altogether, especially in low-mass halos and at low redshift, but that the impact of outflows is suppressed with increasing halo mass.  
By resolving the accretion rates versus radius from the halo centers, we show how cold streams penetrate the hot atmospheres of massive halos at $z\geq2$, but gradually disappear at lower redshift. 
The total baryon mass fraction is also strongly suppressed by outflows in low-mass halos, but is nearly universal in the absence of feedback in halos above the UV background suppression scale, corresponding to circular velocities $v_{\rm c}\sim 50$ km s$^{-1}$. 
The transition halo mass, at which the gas mass in halos is equal for the cold and hot components, is roughly constant at $\sim10^{11.5}$ M$_{\odot}$ and does not depend sensitively on the wind prescription.  
We provide simple fitting formulae for the cold gas accretion rate and the corresponding efficiency with which dark matter channels cold gas into halos in the no-wind case. 
Finally, we show that cold accretion is broadly consistent with driving the bulk of the highly star-forming galaxies observed at $z\sim2$, but that the more intense star formers likely sample the high end of the accretion rate distribution, and may be additionally fueled by a combination of gas recycling, gas re-accretion, hot mode cooling, and mergers.
\end{abstract}

\section{INTRODUCTION}
Both observations and theoretical models indicate that the accretion of gas
from the intergalactic medium (IGM) is a fundamental driver of galaxy
formation and evolution. Such accretion is necessary to explain the
observed evolution of cold gas in and around galaxies \citep[e.g.,][]{2009ApJ...696.1543P}.
Measurements of the cold gas reservoirs and of gas consumption time scales
also imply that a continuous supply of gas from the IGM is required to
maintain star formation over most of the Hubble time \citep[e.g.,][]{2008ApJ...674..151E, 2010ApJ...717..323B}.  At the same time, theoretical models of galaxy formation in a $\Lambda$CDM universe naturally predict continuous accretion of gas into galaxies \citep[e.g.,][]{2002ApJ...571....1M, 2003ASSL..281..185K, 2003MNRAS.339..312S, 2005MNRAS.363....2K, 2009MNRAS.395..160K, 2009Natur.457..451D, 2011MNRAS.414.2458V}. 
The rates at which halos accrete dark matter have now been well quantified across cosmic time using $N-$body simulations, as a function of halo mass and environment, as well as separated according to whether the accretion is smooth or in the form of mergers \citep[e.g.,][]{2002ApJ...568...52W, 2008MNRAS.383..615N, 2008MNRAS.386..577F, 2009MNRAS.394.1825F, 2010MNRAS.401.2245F, 2010MNRAS.406.2267F, 2009MNRAS.398.1858M, 2010ApJ...719..229G}. 
The recent progress on this front was possible thanks to the large dynamic range provided by $N-$body simulations such the Millennium and Millennium II simulations \citep[][]{2005Natur.435..629S, 2009MNRAS.398.1150B}. 
The statistics measured from those dark matter simulations, and the derived fitting formulae, are the basis for a wide variety of studies of galaxy formation and evolution \citep[e.g.,][]{2006MNRAS.365...11C, 2006MNRAS.370..645B, 2008MNRAS.391..481S, 2009Natur.457..451D, 2010ApJ...718.1001B, 2010arXiv1008.1786B, 2010ApJ...724..915H}. 
They also complement basic analytic tools like the Press-Schechter and excursion set formalisms \citep[][]{1974ApJ...187..425P, 1991ApJ...379..440B}.

Baryons, especially in gaseous form, are affected by a wealth of additional physical processes, including pressure forces, radiative heating and cooling, shocking, and star formation. 
It is therefore not surprising that they display a much richer phenomenology. 
Understanding the behavior of baryons is at least as fundamental as understanding the underlying dark matter, as they shape the galaxies that we observe. 
Further impetus for systematically quantifying \emph{how} baryons assemble into dark matter halos is provided by recent work on the character of the accretion flows that fuel galaxies. 
Numerical simulations in fact support a picture in which galaxies acquire their gas through two distinct channels, the ``cold'' and ``hot'' modes \citep[][]{2003ASSL..281..185K, 2005MNRAS.363....2K, 2009MNRAS.395..160K, 2003MNRAS.345..349B, 2006MNRAS.368....2D, 2009Natur.457..451D}. 
While this bimodality is closely related to the classical ``rapid cooling'' and ``slow cooling'' regimes \citep[][]{1977MNRAS.179..541R, 1977ApJ...211..638S, 1978MNRAS.183..341W, 1991ApJ...379...52W}, recent studies have shed new light on the properties of the flows and the transition between the two modes. 

According to the simulations, cold accretion (which we define as the mode in which the gas maintains a temperature $T<2.5\times10^{5}$ K deep within the halo) is the main driver of galaxy formation in the sense that most of the baryons that make it into galaxies are accreted via this mode. 
At high redshift, the cold mode proceeds distinctly via filamentary streams. 
The existence of the cold mode and its importance in driving galaxy formation (in analogy with the rapid cooling regime) has been confirmed in analytic studies and in 1D simulations \citep{1997astro.ph.12204F, 2003MNRAS.345..349B, 2006MNRAS.368....2D}, in addition to cosmological simulations by different groups using both smooth particle hydrodynamics (SPH) and grid-based codes \cite[e.g.][]{2008MNRAS.390.1326O, 2009ApJ...694..396B, 2011MNRAS.414.2458V}. 
It is therefore a robust theoretical prediction. 
The hot mode, in which the accreting gas shock heats to roughly the virial temperature before cooling in a more spherically symmetric fashion onto galaxies, is predicted to compete with the cold mode only around the transition halo mass $M_{\rm h}\sim3\times10^{11}$ M$_{\odot}$ at $z\lesssim1$ \citep[][]{2009MNRAS.395..160K}. 

As the cold streams penetrate the halos on a free fall time, they are very effective at transporting gas to their centers and may be connected to a variety of observed phenomena, including high-redshift star-forming galaxies (Elmegreen \& Elmegreen 2006; Genzel et al. 2006; Dekel et al. 2009a,b\nocite{2006ApJ...650..644E, 2006Natur.442..786G, 2009ApJ...703..785D, 2009Natur.457..451D}), \Lya~blobs \citep[e.g.,][but see Faucher-Gigu\`ere et al. 2010\nocite{2010ApJ...725..633F}]{2001ApJ...562..605F, 2009MNRAS.400.1109D, 2010MNRAS.407..613G}, dense absorption systems in quasar and galaxy spectra \citep[e.g.,][]{2009Natur.457..451D, 2011MNRAS.tmpL.208F, 2011MNRAS.413L..51K, 2011ApJ...735L...1S}, and high-velocity clouds around local galaxies \citep[e.g.,][]{2009ApJ...700L...1K}. 
The morphological distinction between the cold and hot modes is also directly relevant to the efficiency of feedback processes in galaxy formation, as the physical requirements for preventing the infall of dense, cold streams or clumps are different from the requirements to prevent a spherical atmosphere from cooling. 

While previous work has investigated the detailed physics of gas accretion \citep[e.g.,][]{2005MNRAS.363....2K, 2009MNRAS.395..160K, 2008MNRAS.390.1326O}, with some broad considerations of the accretion rates into halos and galaxies, no previous study has systematically quantified the baryonic accretion rates a function of halo mass and redshift, for the different components, in a manner analogous to the growth and merger rates for the dark matter. 
As a consequence, theoretical analyses are often based on simplifying assumptions, such as assuming that the gaseous accretion or star formation rates follow the dark matter with a constant proportionality factor over some relevant mass range \citep[e.g.,][]{2009Natur.457..451D, 2010ApJ...718.1001B, 2010ApJ...717..323B, 2010ApJ...724..895K, 2010arXiv1002.3257C}. 
An important exception is the recent study of \cite{2011MNRAS.414.2458V}, who quantified many of the relevant quantities using hydrodynamical simulations from the OWLS project \citep[][]{2010MNRAS.402.1536S} in a general study of gas accretion onto galaxies and halos. 
Our focus here is however complementary in that we extend the general studies of \cite{2005MNRAS.363....2K} and \cite{2009MNRAS.395..160K}, using a new set of higher resolution simulations and with different outflow parameterizations, to specifically quantify the accuracy of common approximations based on dark matter dynamics. 
Our study of halo accretion is also more detailed than that of \cite{2011MNRAS.414.2458V} in certain respects, as we explore the dependence on the reference radius, cover a wider redshift interval, and explicitly quantify the accuracy of several common scaling relations. 
Our simulations have a mass resolution better by a factor of four than the ones analyzed by \cite{2010MNRAS.406.2325O}, who investigated stellar mass growth in hydrodynamical simulations but did not focus on the accretion rates.

\begin{table*}
\centering
\caption{Hydrodynamical Simulations\label{simulations table}}
\begin{tabular}{|ccccccc|}
\hline\hline
Name                                 & $L$ ($h^{-1}$ Mpc)\tablenotemark{a}                                     & $N$\tablenotemark{b} & $\epsilon$ ($h^{-1}$ kpc)\tablenotemark{c} & $\eta$\tablenotemark{d}      & $\chi$\tablenotemark{e}      & $v_{\rm w}$ (km s$^{-1}$)\tablenotemark{f}      \\
\hline
gdm40n512      & 40 & $2\times512^{3}$ & 1.6 & 0 & 0 & 0 \\ 
gdm40n512\_winds      & 40 & $2\times512^{3}$ & 1.6 &1 & 0.25 & 342 \\
gdm40n512\_swinds      & 40 & $2\times512^{3}$ & 1.6 & 2 & 0.5 & 342 \\
gdm40n512\_fwinds      & 40 & $2\times512^{3}$ & 1.6 & 2 & 2 & 684 \\
\hline
\tablenotetext{a}{Comoving box side length.}
\tablenotetext{b}{Total number of dark matter+gas particles in the box.}
\tablenotetext{c}{Comoving Plummer equivalent gravitational softening length.}
\tablenotetext{d}{Wind mass loading factor.}
\tablenotetext{e}{Fraction of supernova kinetic energy injected into winds.}
\tablenotetext{f}{Wind velocity (eq. (\ref{wind velocity})).}
\tablecomments{Dark matter, gas, and stellar particle masses are $3\times10^{7}$ $h^{-1}$ M$_{\odot}$, $6\times10^{6}$ $h^{-1}$ M$_{\odot}$, and $3\times10^{6}$ $h^{-1}$ M$_{\odot}$, respectively.}
\end{tabular}
\end{table*}

The richer physics of gas accretion, relative to the dark matter, introduces a complication: the accretion rates can be quite sensitive to feedback, especially in the form of galactic winds. 
Outflows must be considered in galaxy formation models, as they are directly observed \citep[e.g.,][]{2000ApJS..129..493H, 2001ApJ...554..981P, 2003ApJ...588...65S, 2009ApJ...703.1394M, 2010ApJ...717..289S}, expected theoretically \citep[e.g.,][]{2000MNRAS.317..697E, 2005ApJ...618..569M, 2011ApJ...735...66M}, and inferred from comparison of the measured star formation history and stellar mass density with cosmological simulations (Springel \& Hernquist 2003; Kere\v{s} et al. 2009b; Schaye et al. 2010; Oppenheimer et al. 2010)\nocite{2003MNRAS.339..312S, 2009MNRAS.396.2332K, 2010MNRAS.402.1536S, 2010MNRAS.406.2325O}.
Unfortunately, the exact nature and properties of feedback remain major unsolved problems in galaxy formation, so that accretion rates measured from hydrodynamical simulations risk being model dependent.  
In this paper, we demonstrate the dependence on the outflow parameters using a suite of otherwise identical simulations. 
We show that, even in the absence of outflows, the cold gas accretion rate is in general not a universal (or constant) fraction of the dark matter accretion rate. 
In the presence of outflows, the cold gas accretion rate itself can decouple from the star formation rate within the halo. 
In general, the star formation rate within low-mass halos is suppressed with respect to the cold gas accretion rate because the gas can be ejected and unbound from the halo before forming stars. 
Additionally, the energy and momentum injected in the halo by outflows can in some regimes directly slow down gas accretion. 
Sufficiently massive halos can however retain the gas ejected from the galaxies they contain, and re-accretion can boost the star formation rates in those halos with respect to the no-feedback case \citep[e.g.,][]{2010MNRAS.406.2325O}. 
In addition to the accretion rates, we also quantify the baryon mass fraction within halos, and show that it is also strongly suppressed by outflows in low-mass halos. 
Our simulation without outflows resolves the mass scale at which photoionization by the UV background suppresses the total baryon fraction at $z\lesssim 5$, corresponding to a circular velocity $v_{\rm c}\sim50$ km s$^{-1}$. 
We provide simple fitting formulae for the cold gas accretion rate as a function of halo mass and redshift in the no-wind case, and compare it to recent star formation rate estimates of $z\sim2$ galaxies. 
Our results should be useful to improve the accuracy of analytic \citep[e.g.,][]{2010ApJ...718.1001B}, semi-analytic \citep[e.g.,][]{2006MNRAS.370.1651C, 2009ApJ...700L..21K, 2011MNRAS.tmp.1035L, 2010MNRAS.405.1690D, 2011MNRAS.410.2653B}, and empirical \citep[e.g.,][]{2009ApJ...696..620C} models of galaxy formation. 

The plan of this paper is as follows. 
In \S \ref{simulations}, we describe our numerical simulations and methodology. 
The accretion rates and baryon fractions measured from our suite of simulations are presented in \S \ref{numerical results}, where we also discuss the implications of our results for dark matter-based scalings and provide relevant fitting formulae. 
We discuss our results and conclude in \S \ref{discussion}.

\section{SIMULATIONS AND METHODOLOGY}
\label{simulations}

\subsection{Code Details}
\label{code details}
Our cosmological hydrodynamical simulations, run with the GADGET code \citep[][]{2005MNRAS.364.1105S}, are similar to the ones presented in \cite{2009MNRAS.395..160K} and \cite{2010ApJ...725..633F} (the \verb|gdm40n512| simulation analyzed here is the same as in the latter article). 
Gravitational forces are calculated using a combination of the particle mesh algorithm \citep[e.g.,][]{1988csup.book.....H} for large separations and the hierarchical tree algorithm \citep[e.g.,][]{1986Natur.324..446B, 1987ApJS...64..715H} at small distances.
The gas dynamics is computed using a smoothed particle hydrodynamics (SPH) algorithm \citep[e.g.,][]{1977AJ.....82.1013L, 1977MNRAS.181..375G} that conserves both energy and entropy \citep[][]{2002MNRAS.333..649S}.

The modifications with respect to the public version of the code include the treatment of cooling, the effects of a uniform ultra-violet (UV) background, and a multiphase star formation algorithm as in \citet[][]{2003MNRAS.339..289S}.
Star formation is implemented by the stochastic spawning of collisionless star particles by the gas particles.
In practice, star formation in the multiphase model occurs above a density threshold of $n_{\rm H}=0.13$ cm$^{-3}$ and is calibrated to the observed \cite{1998ApJ...498..541K} law. 
The thermal and ionization properties of the gas are calculated including all the relevant processes in a plasma with primordial abundances of hydrogen and helium following \cite{1996ApJS..105...19K}. 
We use the UV background model of \cite{2009ApJ...703.1416F}, empirically calibrated to match IGM absorption measurements (Faucher-Gigu\`ere 2008a,b,c\nocite{2008ApJ...682L...9F, 2008ApJ...688...85F, 2008ApJ...681..831F}).  

\begin{figure*}
\begin{center} 
\includegraphics[width=1.0\textwidth]{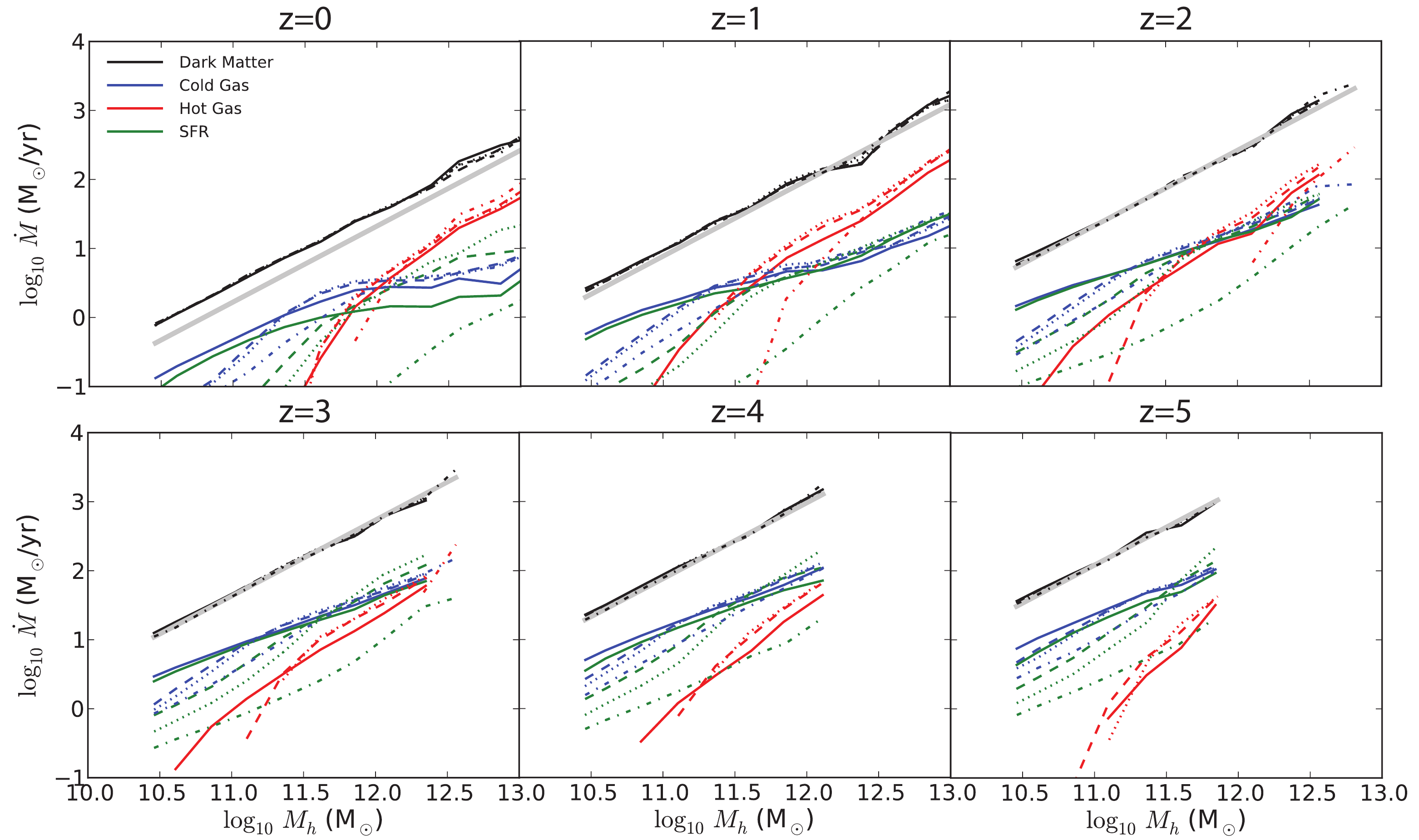}
\end{center}
\caption[]{Comparison of the median accretion rates through the virial shell for the different wind prescriptions. 
\emph{Solid curves}: no winds. 
\emph{Dashed:} constant-velocity winds with $v_{\rm w}=342$ km s$^{-1}$ and mass loading $\eta=1$ ({\tt winds}). 
\emph{Dotted:} constant-velocity winds with $v_{\rm w}=342$ km s$^{-1}$ and mass loading $\eta=2$ ({\tt swinds}).
\emph{Dash-dotted:} constant-velocity winds with $v_{\rm w}=684$ km s$^{-1}$ and mass loading $\eta=2$ ({\tt fwinds}).
The thick grey lines show the fit of FMB10 to the dark matter growth rate, scaled by $(\Omega_{\rm m}-\Omega_{\rm b})/\Omega_{\rm m}$ since these authors studied a dark matter-only simulation. 
}
\label{compare wind sims} 
\end{figure*}

\subsection{Galactic Winds and Simulation Parameters}
\label{galactic winds}
To investigate the effects of feedback in the form of galactic winds on the baryonic accretion rates, we use a simple constant-velocity wind model, implemented as in \cite{2003MNRAS.339..289S}. 
Exploring these effects is important, as we have outlined in the introduction, since outflows are known to be ubiquitous in the Universe (especially at high redshift), and to have important implications for the predictions of numerical simulations. 
Our goal here is to show phenomenological examples of how galactic winds can affect the accretion rates of the different baryonic components. 
In particular, we do not imply that constant-velocity winds are an accurate representation of actual galactic winds, nor do we attempt to match the details of observations.
Instead, we simply explore variations of the model parameters to determine features that depend on (or are robust to) the wind properties. 
In future work, we will use more physical implementations of galactic outflows to study their dynamical interactions with accreting material in detail.

As described in \cite{2003MNRAS.339..289S}, the wind model is parameterized by a constant mass loading factor $\eta$ and a constant fraction of supernova energy deposited into the wind $\chi$. Specifically, the disk mass-loss rate in the wind is defined as $\dot{M}_{\rm w} = \eta \dot{M}_{\star}$, where $\dot{M}_{\star} \equiv SFR$ is the star formation rate, and the velocity of the wind as it leaves the disc satisfies the equation
\begin{equation}
\frac{1}{2}\dot{M}_{\rm w} v_{\rm w}^{2} = \chi \epsilon_{\rm SN} \dot{M}_{\star},
\end{equation}
i.e.,
\begin{equation}
\label{wind velocity}
v_{\rm w} = \sqrt{
\frac{2 \chi \epsilon_{\rm SN}}{\eta}
}.
\end{equation}
In the equations above, $\epsilon_{\rm SN}=4.6\times10^{48}$ erg M$_{\odot}^{-1}$ is the average energy output by supernovae per unit stellar mass, corresponding to the canonical value of $10^{51}$ erg for a Salpeter initial mass function (IMF) between 0.1 and 40 M$_{\odot}$, assuming that stars with $>8$ M$_{\odot}$ die as supernovae. 
Gas particles are stochastically incorporated into the wind according to their star formation rate and the wind particles are briefly decoupled from hydrodynamic interactions. 
This approach has been shown to yield numerically converged results even at the relatively coarse resolution of cosmological simulations \citep[][]{2003MNRAS.339..289S}. 
In order to produce roughly bipolar outflows normal to galactic discs, the wind particles are given initial velocities parallel to $\pm \nabla \phi \times {\bf v}$, where $\phi$ is the local gravitational potential and ${\bf v}$ is the velocity of the star-forming particle sourcing the wind. 

Our simulations implement a basic form of instantaneous gas recycling from Type II supernovae, in which a fraction $\beta_{\rm SN}=0.1$ of the star formation rate is immediately returned as interstellar gas. 
Simulated galaxies can therefore maintain steady-state star formation rates exceeding the external gas supply rate by a fraction $\beta_{\rm SN}$. 
The additional star formation however does not contribute to the net rate of stellar mass growth and the parameter $\beta_{\rm SN}$ is sensitive to uncertain IMF and stellar evolution assumptions.
In order to subtract out the model dependence on $\beta_{\rm SN}$, we henceforth make the identification $SFR \to SFR^{\rm tot}(1-\beta_{\rm SN})$ unless otherwise specified. 
I.e., the $SFR$s we report do not include the boost from gas return. 
This convention has the benefit that $SFR$s exceeding the cosmological supply rate are directly indicative of additional supply channels, e.g. via wind fall-back (either on the original galaxy, or on a neighbor) or hot gas cooling. 
To be exact, our $SFR$s should however be multiplied by $(1-\beta_{\rm SN})^{-1}\approx1.1$ when comparing with measured star formation rates. 

The simulations analyzed in this work are detailed in Table \ref{simulations table}. 
The cosmological parameters for all these simulations are the best-fit parameters for the WMAP5+BAO+SN data \citep[][]{2009ApJS..180..330K}: $(\Omega_{\rm m},~\Omega_{\rm b},~\Omega_{\Lambda},~h,~\sigma_{8},~n_{\rm s})=(0.28,~0.046,~0.72,~0.7,~0.82,~0.96)$. 
All simulations have a box side length of 40~$h^{-1}$ comoving Mpc and use $512^{3}$ particles to represent each of the dark matter and gas components. 
They assume a \cite{1999ApJ...511....5E} transfer function and the $N-$body integration begins at redshift $z=99$. 
 The dark matter, gas, and stellar particle masses are $3\times10^{7}$ $h^{-1}$ M$_{\odot}$, $6\times10^{6}$ $h^{-1}$ M$_{\odot}$, and $3\times10^{6}$ $h^{-1}$ M$_{\odot}$, respectively. 

The basic simulation \verb|gdm40n512| does not include feedback, except for a sub-resolution multiphase model to pressurize the interstellar medium (ISM) \citep[][]{2003MNRAS.339..289S}. 
Three otherwise identical simulations incorporate galactic winds as described above, with parameters given in the table. 
Briefly, the winds in the \verb|winds| and \verb|swinds| runs have a launch velocity of $v_{\rm w}=342$ km s$^{-1}$, and mass loading factors $\eta=1$ and $\eta=2$, respectively. 
The \verb|swinds| model is similar to the slow winds (sw) of \cite{2010MNRAS.406.2325O}, which they find produce agreement with the observed stellar mass fraction at $z=0$ comparable to their best momentum-driven winds (though not as good at the low-mass end).
\footnote{The simulations of \cite{2010MNRAS.406.2325O} differ slightly by their inclusion of metal cooling and asymptotic giant branch (AGB) gas recycling, but we have verified that our {\tt swinds} simulation produces a consistent mass function at $z=0$.} 
The third wind simulation (\verb|fwinds|) uses an extreme set of parameters, $\eta=2$ and 
$\chi=2$, which corresponds to driving the winds with twice as much kinetic energy as supernovae fiducially provide, and ejecting twice as much mass in the winds as the star formation rate. 
The wind velocity in this case is 684 km s$^{-1}$. 
Although other feedback mechanisms, for example radiation pressure from massive stars \citep[e.g.,][]{2005ApJ...618..569M}, can accelerate winds with velocities or mass loadings greater than possible for energy injection by supernovae alone, we have verified that these parameters yield a baryon mass function at $z=0$ heavily suppressed relative to observations \citep[e.g.,][]{2003ApJS..149..289B} for 2 decades in galaxy mass around $L^{\star}$ (similar to the cw model of Oppenheimer et al. 2010\nocite{2010MNRAS.406.2325O}). 

The simulations presented in this work have a mass resolution $11\times$ finer than the main simulation analyzed by \cite{2009MNRAS.395..160K}, in a comparable volume ($2\times288^{3}$ particles with 50 $h^{-1}$ comoving Mpc box side length vs. $2\times512^{3}$ particles with a 40 $h^{-1}$ comoving Mpc box side length), and use the same code. 
Our finer mass resolution is also slightly better than that of the smaller 11 $h^{-1}$ comoving Mpc box (with $2\times128$ particles) used by \cite{2009MNRAS.395..160K} to test the numerical convergence of their results, so that we can be confident that our findings, broadly, are robust. 
It is worth noting that many of the quantities that we analyze, being evaluated at the virial radius and therefore involving lower densities, should actually have significantly less stringent convergence requirements than the accretion rates onto galaxies studied by \cite{2005MNRAS.363....2K} and \cite{2009MNRAS.395..160K}. 
At the detailed quantitative level, it is however possible that even the accretion rates through the virial shell could be subject to minor resolution effects. 
Two effects stand out in particular. 
First, a higher resolution allows the formation of lower-mass objects, which can remove gas by locking baryons into stars. 
Because structure formation proceeds hierarchically, the effect can propagate to higher-mass halos. 
\cite{2009MNRAS.395..160K} saw this effect at the 30\% level on the \emph{smooth} gas accretion rates. 
This effect should be significantly smaller in our simulations for several reasons. 
Since we do not explicitly distinguish between smooth and clumpy accretion, it is only the fraction of baryons locked into stars in low-mass halos that can affect our results. 
As our simulations resolve the UV background suppression scale at redshifts $z\lesssim5$ (\S \ref{baryonic contents of halos}), there is only a small regime of halo mass and redshift for which increasing the resolution might yield more star formation in low-mass halos. 
Furthermore, at $z\geq5$ and for the minimum halo mass $M_{\rm h}\approx10^{10.5}$ M$_{\odot}$ at which we report accretion rates, the accretion rate of stars in our no-wind simulation is only $<3\%$ of the cold gas accretion rate. 
The convergence requirements for the simulations with outflows are less severe in this respect because outflows are effective at removing baryons from low-mass halos. 
A second potential effect concerns high-mass halos, in which gas from a hot atmosphere can cool onto galaxies and also fuel star formation. 
This may in fact be somewhat sensitive to resolution, but in this case the situation is less clear and different numerical methods have not yet converged to a common answer \citep[e.g.,][]{2009MNRAS.395..160K}. 
This issue will be explored by comparing GADGET results with the moving-mesh code AREPO in a subsequent paper (Kere\v{s} et al., in prep.)\nocite{keres_arepo}.

\begin{figure}
\begin{center} 
\includegraphics[width=0.475\textwidth]{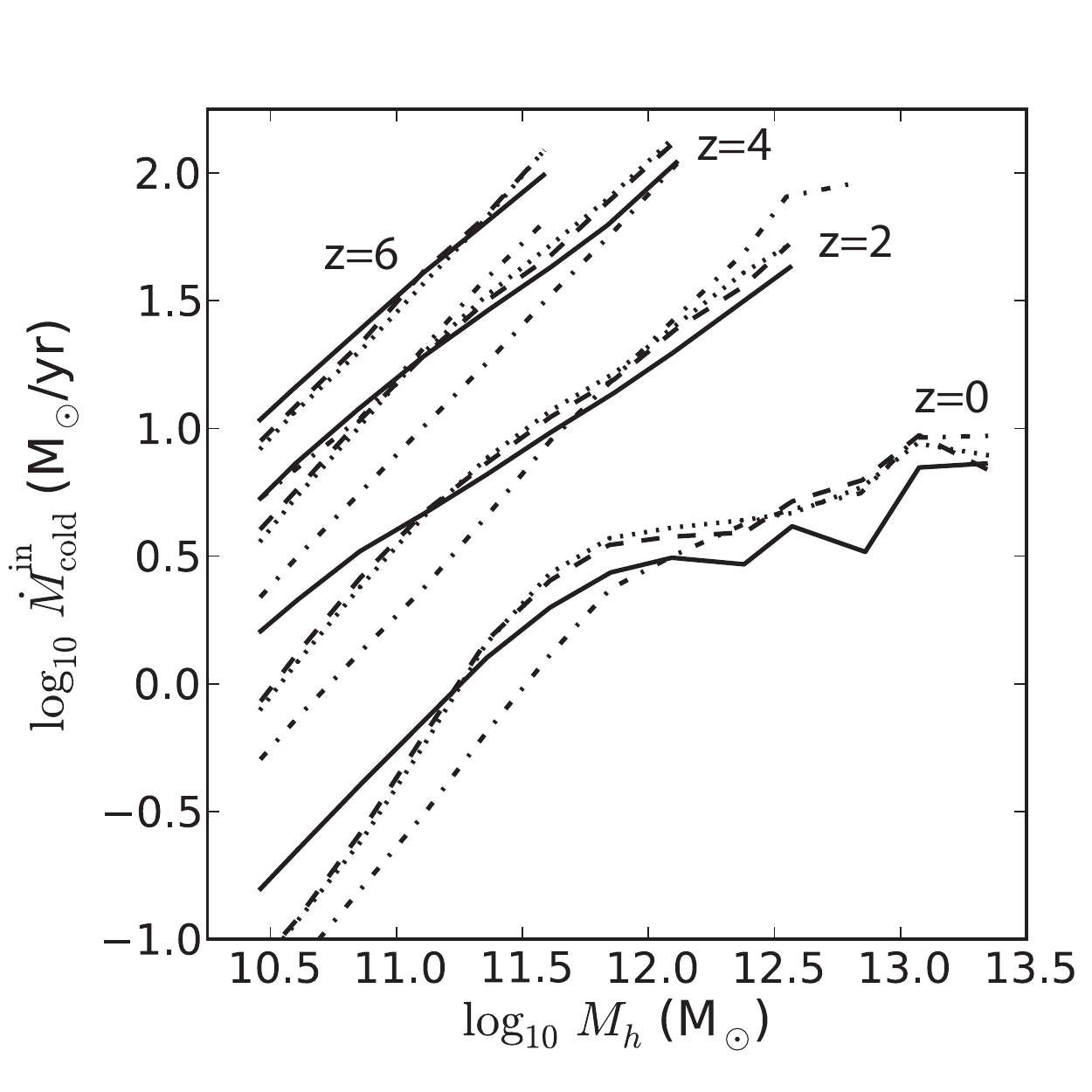}
\end{center} 
\caption[]{Comparison of the cold gas accretion rates through the virial shell for the different wind prescriptions. Only the gas particles with inward radial motion are included in order to exclude the negative contributions from outflowing material. 
From the bottom up, the results are shown at $z=0, 2, 4,$ and 6. 
\emph{Solid curves}: no winds. 
\emph{Dashed:} constant-velocity winds with $v_{\rm w}=342$ km s$^{-1}$ and mass loading $\eta=1$ ({\tt winds}). 
\emph{Dotted:} constant-velocity winds with $v_{\rm w}=342$ km s$^{-1}$ and mass loading $\eta=2$ ({\tt swinds}).
\emph{Dash-dotted:} constant-velocity winds with $v_{\rm w}=684$ km s$^{-1}$ and mass loading $\eta=2$ ({\tt fwinds}).}
\label{Mdot cold in comparison} 
\end{figure}

\begin{figure}
\begin{center} 
\includegraphics[width=0.475\textwidth]{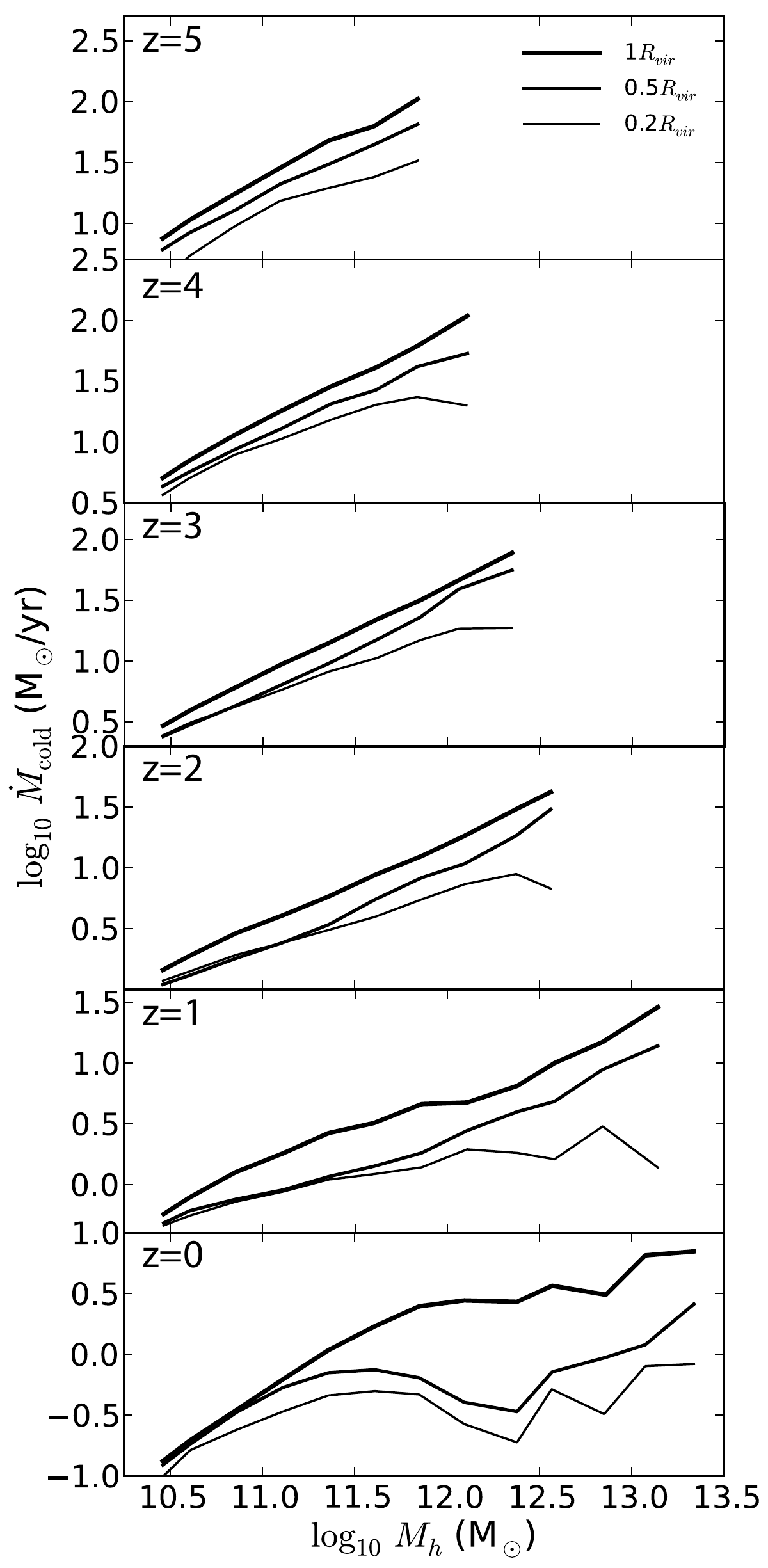}
\end{center} 
\caption[]{Net cold gas accretion rate as a function of redshift and halo mass, for the simulation without outflows, through shells of radii 1$R_{\rm vir}$, 0.5$R_{\rm vir}$, and 0.2$R_{\rm vir}$. 
The panels show how deep the cold gas penetrates into halos in different regimes. The vertical axes cover exactly 2 orders of magnitude in each panel, so that relative vertical separations in the different panels are fair representations of ratios.
}
\label{Mdot cold vs r} 
\end{figure}

\subsection{Halo Identification}
\label{halo identification}
A friends-of-friends (FoF) algorithm \cite[e.g.,][]{1985ApJ...292..371D} with linking length set to $b=0.2$ in units of the mean interparticle separation is used to identify the dark matter halos in the simulations.
The total mass of the particles within each FoF group, $M_{\rm FoF}$, corresponds approximately to the mass in a sphere of mean interior density 180 times the background matter density, $M_{180b}$. 
\cite{2002ApJS..143..241W} in fact showed that $M_{180b}$ is a more accurate proxy for the FoF mass than the common virial mass definition of 200 times the critical density. 
We therefore define the virial radius of a halo as the radius of a sphere containing $M_{180b}$, $R_{\rm vir}\equiv R_{180b} \approx [M_{\rm FoF} / 240 \pi \rho_{u}(z)]^{1/3}$, where $\rho_{u}(z)=\rho_{\rm crit}\Omega_{m}(1+z)^{3}$ and $\rho_{\rm crit}$ is the critical density at $z=0$.
In what follows, we use $M_{\rm h}$ as a shorthand for $M_{\rm FoF}\approx M_{180b}$. 
The center of each halo is defined as the point deepest in the gravitational potential.
Since we run the FoF algorithm on the dark matter particles only, we multiply the returned masses by $\Omega_{\rm m}/(\Omega_{\rm m}-\Omega_{\rm b})$ to account for the baryons. 
The use of the FoF mass is consistent with the previous work of \cite{2009MNRAS.398.1858M} and \cite{2010MNRAS.406.2267F} on the dark matter mass assembly, and also allows for the direct use of existing analytic approximations to the halo mass function (such as that of Seth \& Tormen 2002\nocite{2002MNRAS.329...61S}, which is calibrated to FoF masses) in applications.

\subsection{Definition of Accretion Rates}
\label{definition of accretion rates}
Our approach to calculating the accretion rates differs from the technique often used in Lagrangian simulations (either dark matter-only or SPH) of identifying particles belonging to halos or galaxies at different redshifts and dividing the mass difference by the time interval \citep[e.g.,][]{2009MNRAS.398.1858M, 2010MNRAS.406.2267F, 2002ApJ...571....1M, 2006ApJ...647..763M, 2009MNRAS.395..160K, 2011MNRAS.414.2458V}. 
Since the thermal state of the gas, in particular whether it is classified as ``hot'' or ``cold'', can be a strong function of radius from the halo center, it is advantageous to measure the accretion rates through shells of prescribed radii. 
This allows us to investigate how the mix of different baryonic components evolves during the infall. 
Furthermore, this approach allows us to select particles based on the direction of their motion, which as we will see is useful for separating accreting material from outflows. 
Mass fluxes also have the significant advantage that they can be identically computed in grid-based simulations \citep[e.g.,][]{2008MNRAS.390.1326O}; results from this work can therefore be directly compared with those obtained with grid codes, including the new moving-mesh code AREPO \citep[][]{2010MNRAS.401..791S}. 
Numerical noise from the finite number of particles overlapping with a relatively thin shell is mitigated by evaluating the accretion rates only for halos containing more than 500 dark matter particles and, on average, an equal number of gas particles. Furthermore, all of our results regarding low-mass halos are aggregate statistics over large samples and so are robust. 
In this work, we do not explicitly distinguish between mergers and smooth accretion, but we do consider gas that is part of the ISM separately from the rest (see below), which we subsequently refer to as ``diffuse'', as explained below.

Since the natural scale at which to evaluate the accretion rates is the virial radius, we will usually refer to the ``virial shell'', although in general we can set its radius $R_{\rm s}(M,~z)$ to any factor of the virial radius. 
The thickness of the shell is labeled by $\Delta R_{\rm s}(M,~z)$, and it is centered on the point deepest in the halo potential. 
The net accretion rate of a given component $i$ is then defined as 
\begin{equation}
\label{accretion rate sum}
\dot{M}_{i} = \sum_{\rm p_{i}} M_{\rm p_{i}} { \frac{{\bf v_{\rm p_{i}}-v_{\rm A}}}{\Delta R_{\rm s}} \cdot \frac{{\bf r}_{\rm h}-{\bf r}_{\rm p_{i}}}{ |{\bf r}_{\rm h}-{\bf r}_{\rm p_{i}}|}},
\end{equation}
where the sum is over the particles of the component of interest that intersect the virial shell, i.e. $R_{\rm s}-\frac{\Delta R_{\rm s}}{2} \leq |{\bf r}_{\rm p_{i}}-{\bf r}_{\rm h}| \leq R_{\rm s}+\frac{\Delta R_{\rm s}}{2}$.
Here, $M_{\rm p_{i}}$ is the particle mass, ${\bf r}_{\rm p_{i}}$ is its position vector, $\bf v_{\rm p_{i}}$ is its velocity vector, and ${\bf r}_{\rm h}$ is the position of the halo center. 
The velocity vector ${\bf v}_{\rm A}$, to be discussed in detail in the next section, accounts for the motion of the area element on virial shell itself. 
The velocity vector $\bf v_{\rm p_{i}}$ includes both the peculiar velocity contribution and a radial component $H(z)R_{\rm s}$ (assumed constant within a given radial shell) accounting for the Hubble expansion. 
Unless otherwise noted, all quantities are in proper units.
Note that the contribution of any particle to the net accretion rate is positive if it is moving inward with respect to the virial shell, and negative otherwise. 
We will also consider the total mass enclosed within the virial shell of different components, defined as
\begin{equation}
M_{i} = \sum_{\rm p_{i}} M_{\rm p_{i}},
\end{equation}
where the sum is over all particles with $|{\bf r}_{\rm p_{i}}-{\bf r}_{\rm h}| \leq R_{\rm s}$. The star formation rates within halos are evaluated similarly. 

We keep track of five species separately: dark matter (DM), cold gas ($n_{\rm H}<0.13$ cm$^{-3}$, $T\leq2.5\times10^{5}$ K), hot gas ($n_{\rm H}<0.13$ cm$^{-3}$, $T>2.5\times10^{5}$ K), stars ($\star$), and ISM ($n_{\rm H}\geq0.13$ cm$^{-3}$). 
The temperature threshold between the cold and hot phases is motivated by the bimodality found in cosmological simulations for the maximum temperature attained by gas accreting from the IGM \citep[e.g.,][]{2005MNRAS.363....2K}, and approximately distinguishes between gas that is shock-heated as it falls in, and gas that is not significantly shocked. 
Note, however, that the cold and hot gas accretion rates through shells considered in the present work differ from the ``cold mode'' and ``hot mode'' defined by \cite{2005MNRAS.363....2K} and \cite{2009MNRAS.395..160K}. 
In those papers, the cold and hot modes are defined in terms of the \emph{maximum} temperature attained before accretion onto galaxies, rather than the instantaneous temperature. 
The ISM material is defined in the context of the \cite{2003MNRAS.339..289S} sub-resolution multiphase star formation model used in our simulations. 
As outlined in \S \ref{code details}, the gas particles with density $n_{\rm H}\geq 0.13$ cm$^{-3}$ are assumed to develop a multiphase structure. 
These multiphase particles carry an effective temperature that is a mass-weighted average of cold and hot ISM phases. 
As a result of the averaging procedure, these effective temperatures can exceed the $T=2.5\times10^{5}$ K threshold between the cold and hot phases, even though most of the gas mass is usually cold. 
We therefore treat the ISM material separately and never include it in the cold and hot components, which therefore always have have densities $n_{\rm H}<0.13$ cm$^{-3}$.
Like cold accretion from the IGM (but unlike hot gas in massive halos), gas that is accreted by a halo in ISM form (e.g., via mergers) can rapidly form stars. 
It will therefore be useful to consider the sum of the cold and ISM accretion rates in some circumstances.  
The net accretion rates are further divided into ``inward'' and ``outward'' parts, such that $\dot{M}_{i} \equiv \dot{M}_{i}^{\rm in}-\dot{M}_{i}^{\rm out}$. The inward rates are defined by summing only over particles with $({\bf v_{\rm p_{i}} - {\bf v_{\rm A}}) \cdot (\bf r}_{\rm h}-{\bf r}_{\rm p_{i}})>0$, as in equation (\ref{accretion rate sum}), and vice versa.
 
Because our Eulerian definitions of the cold and hot gas accretion rates differ from the Lagrangian definitions of the cold and hot modes of \cite{2005MNRAS.363....2K} and \cite{2009MNRAS.395..160K}, some systematic differences are anticipated. 
Since we have adopted the same temperature threshold between the cold and hot components as in these previous papers, but consider the instantaneous instead of maximum temperature, we expect our cold gas accretion rates onto halos to be systematically higher. 
A similar consideration applies when comparing with the Lagrangian definition of \cite{2011MNRAS.414.2458V}, but some deviations from this expectation could arise due to the higher temperature threshold $T=10^{5.5}$ K adopted by those authors. 
Where our results overlap, they are broadly consistent with those previous studies and with the adaptive mesh refinement simulation of \cite{2008MNRAS.390.1326O}, as we will detail where applicable. 
We note, however, that many of our results are new and not directly comparable with earlier SPH analyses. 
For example, \cite{2009MNRAS.395..160K} focused on accretion onto galaxies rather than halos, and \cite{2011MNRAS.414.2458V} did not resolve the accretion as a function of radius or based on whether the gas elements were inflowing or outflowing. 

It is also noteworthy that the simulations of \cite{2011MNRAS.414.2458V} implemented some physics differently from our simulations. 
In particular, their reference model included metal line cooling, which they found to have an important effect on accretion onto galaxies, but a much smaller impact on halo accretion. 
An importance distinction between the cold vs. hot accretion fractions reported herein and in \cite{2009MNRAS.395..160K} is that in the latter work only gas that ended up into galaxies was considered. 
Since the gas accreted by galaxies is strongly biased toward the cold mode, especially in massive halos in which the cooling times for gas shock-heated to the virial temperature are long, the fraction of cold vs. hot mode accretion onto galaxies can be much larger than the relative cold vs. mode hot rates through the virial radius. 
As we will show (\S \ref{SFR Mdot cold section}), the cold gas accretion rates at the virial radius is closely related to the star formation rates within halos, and therefore share many of the key features of the Lagrangian cold mode defined by \cite{2005MNRAS.363....2K} and \cite{2009MNRAS.395..160K}. 

Since our accretion rates are evaluated directly using the velocity vectors of the particles, they are instantaneous. 
On the other hand, accretion rates evaluated by dividing the mass by the time interval between discrete snapshots are implicitly time-averaged. 
While time averaging preserves the mean accretion rates, it in general modifies the shape of the distribution, reducing its scatter, and can therefore affect statistics such as its median. 
This effect will be encountered in \S \ref{accretion rates}, where we compare the median dark matter accretion rates inferred from the Millennium simulations to ours.

Plotting the halo accretion rates and baryon contents as a function of halo mass requires some care, since the slope of the halo mass function implies that the data points in any given mass bin are in general concentrated in its lower end. 
To mitigate this potential source of bias, we use the median $M_{\rm h}$ in each bin as the horizontal coordinate, rather than its geometric center. 

\begin{figure*}
\begin{center} 
\includegraphics[width=1.0\textwidth]{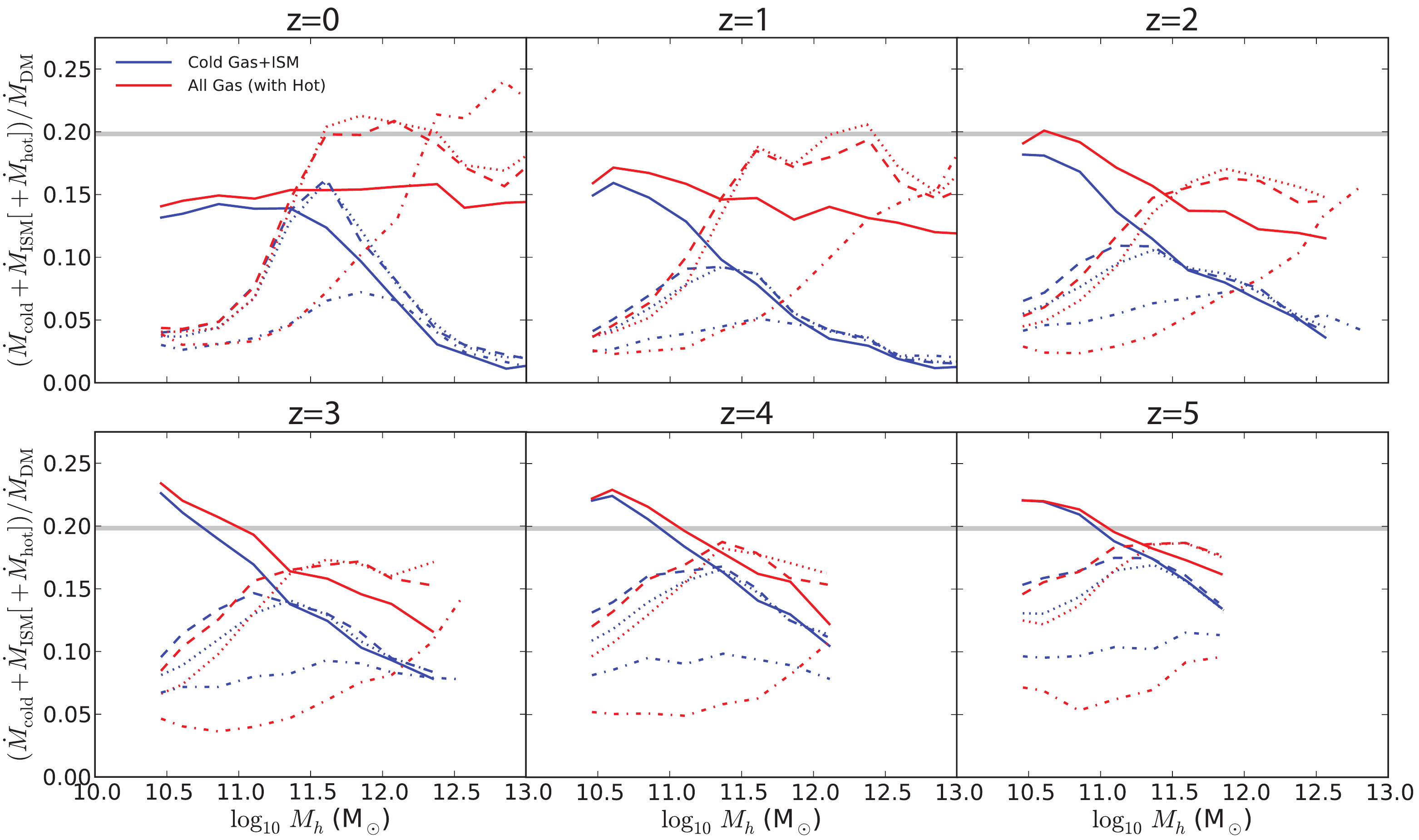}
\end{center} 
\caption[]{Comparison of the ratio of the cold gas+ISM to dark matter accretion rates through the virial shell for the different wind prescriptions (blue). 
The red curves show the same ratio, but including all gas. 
 \emph{Solid curves}: no winds. 
\emph{Dashed:} constant-velocity winds with $v_{\rm w}=342$ km s$^{-1}$ and mass loading $\eta=1$ ({\tt winds}). 
\emph{Dotted:} constant-velocity winds with $v_{\rm w}=342$ km s$^{-1}$ and mass loading $\eta=2$ ({\tt swinds}).
\emph{Dash-dotted:} constant-velocity winds with $v_{\rm w}=684$ km s$^{-1}$ and mass loading $\eta=2$ ({\tt fwinds}).
The thick grey lines show the universal ratio $\Omega_{\rm b}/\Omega_{\rm DM}$. 
}
\label{Mdot cold Mdot DM} 
\end{figure*}

\begin{figure*}
\begin{center} 
\includegraphics[width=1.0\textwidth]{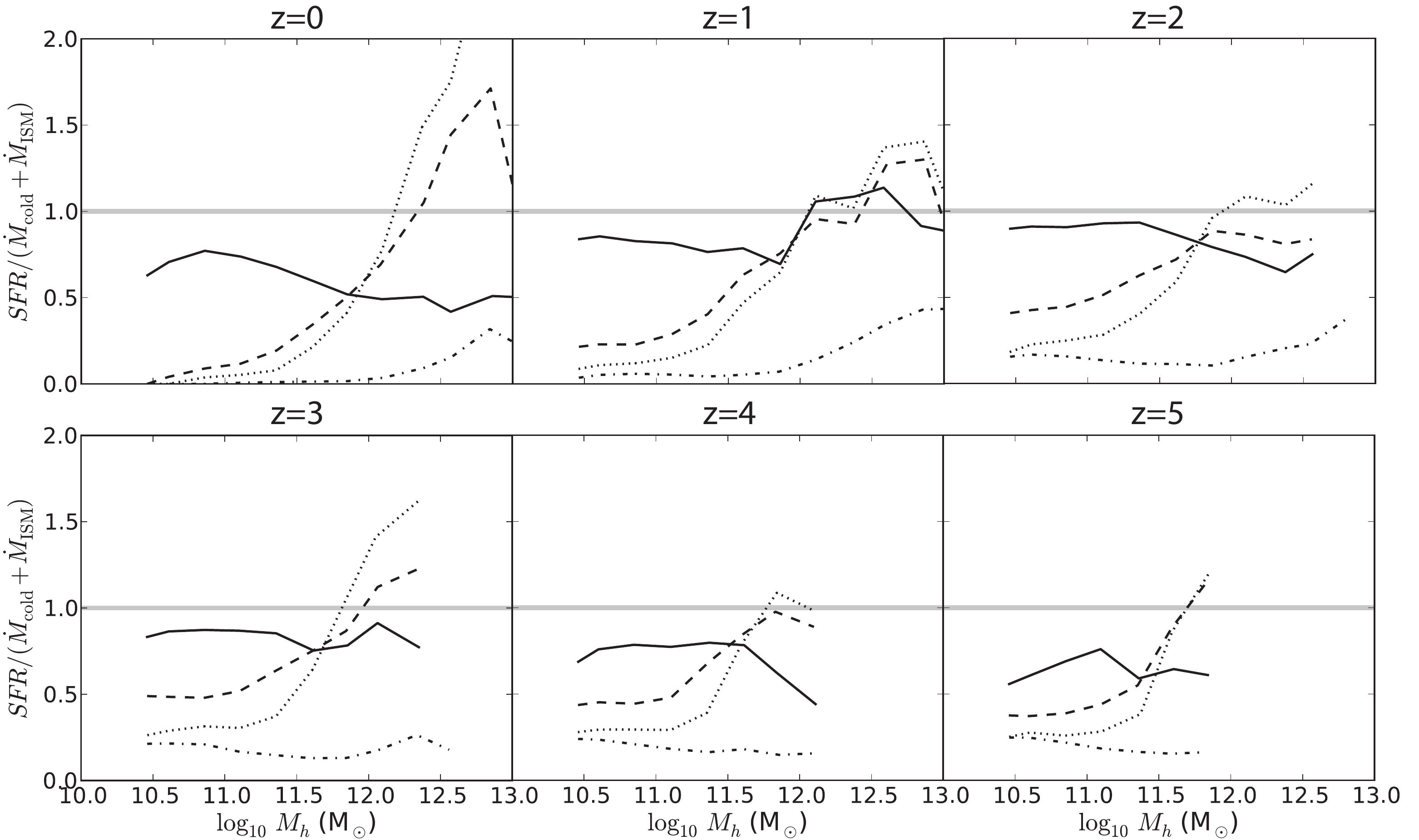}
\end{center}
\caption[]{Comparison of the ratio of the star formation to cold gas+ISM accretion rates through the virial shell for the different wind prescriptions. 
\emph{Solid curves}: no winds. 
\emph{Dashed:} constant-velocity winds with $v_{\rm w}=342$ km s$^{-1}$ and mass loading $\eta=1$ ({\tt winds}). 
\emph{Dotted:} constant-velocity winds with $v_{\rm w}=342$ km s$^{-1}$ and mass loading $\eta=2$ ({\tt swinds}).
\emph{Dash-dotted:} constant-velocity winds with $v_{\rm w}=684$ km s$^{-1}$ and mass loading $\eta=2$ ({\tt fwinds}).
The thick grey lines show the reference value of unity.
}
\label{SFR Mdot cold} 
\end{figure*}

\begin{figure*}
\begin{center} 
\includegraphics[width=1.0\textwidth]{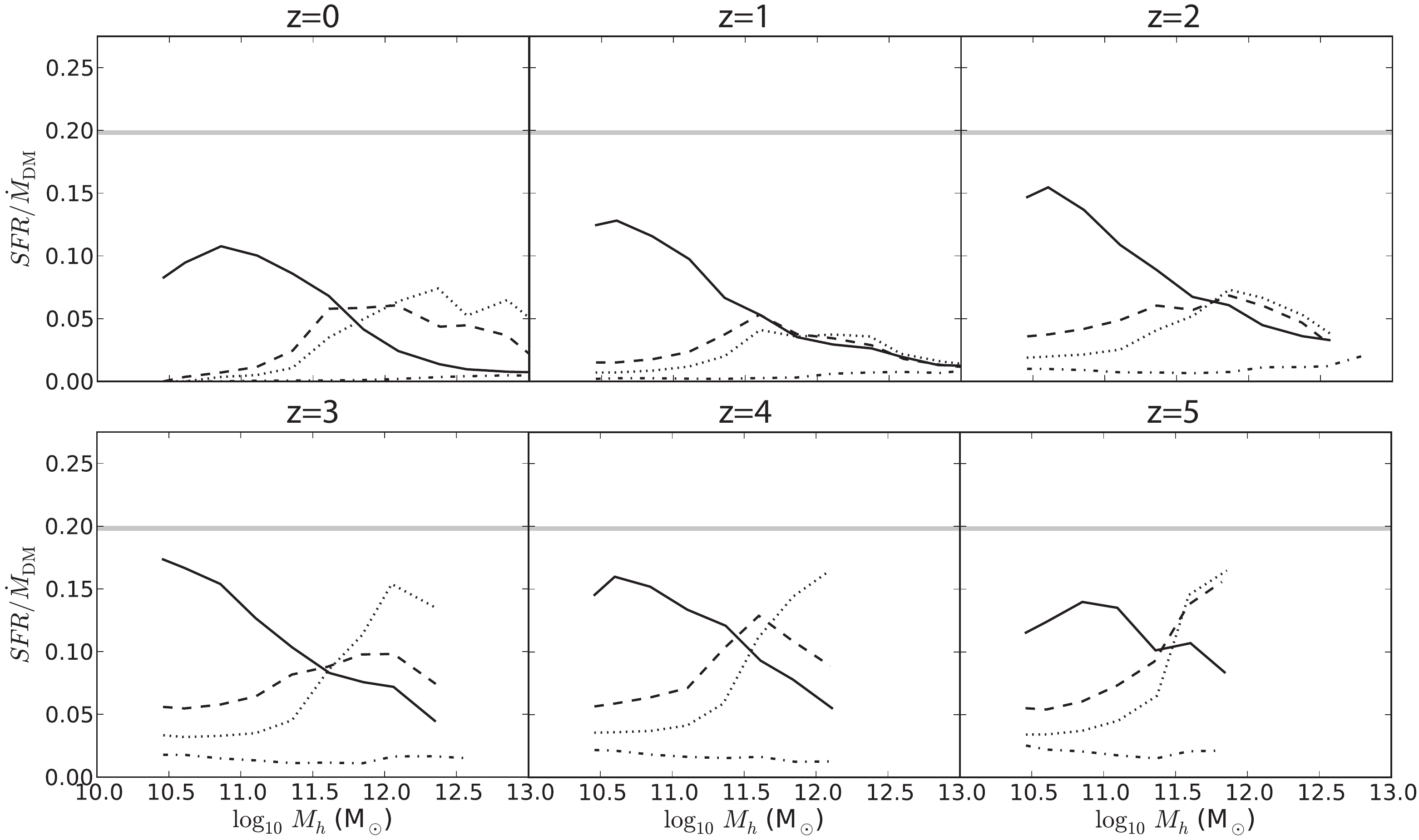}
\end{center}
\caption[]{Comparison of the ratio of the star formation to dark matter accretion rates through the virial shell for the different wind prescriptions. 
\emph{Solid curves}: no winds. 
\emph{Dashed:} constant-velocity winds with $v_{\rm w}=342$ km s$^{-1}$ and mass loading $\eta=1$ ({\tt winds}). 
\emph{Dotted:} constant-velocity winds with $v_{\rm w}=342$ km s$^{-1}$ and mass loading $\eta=2$ ({\tt swinds}).
\emph{Dash-dotted:} constant-velocity winds with $v_{\rm w}=684$ km s$^{-1}$ and mass loading $\eta=2$ ({\tt fwinds}).
The thick grey lines show the universal ratio $\Omega_{\rm b}/\Omega_{\rm DM}$. 
}
\label{SFR Mdot DM} 
\end{figure*}

\subsection{Definition of the Virial Shell}
\label{definition of the virial shell}
The virial shell in general grows with the Hubble expansion and as the halo acquires mass, and the shell also moves with the halo itself. 
An area element on the shell therefore has motion of its own, which must be accounted for in calculating fluxes. 
The velocity vector of a shell area element can be written as
\begin{equation}
{\bf v}_{\rm A} = {\bf v}_{\rm h} + {\bf v}_{\rm exp},
\end{equation}
where the ${\bf v}_{\rm h}$ term accounts for the center-of-mass motion of the halo and the ${\bf v}_{\rm exp}$ term accounts for the expansion of the virial shell.
Let ${\bf r}_{\rm A}$ be the position vector of the area element of interest on the virial shell (assumed to be at the center of its thickness). 
Then, since the shell expansion velocity is proportional to the rate of change of the shell radius,
\begin{equation}
{\bf v}_{\bf exp} = \dot{R}_{\bf s} 
\frac{\bf r_{\rm A} - r_{\rm h}}{\bf |r_{\rm A} - r_{\rm h}|}.
\end{equation}
Although the virial shell can be defined arbitrarily, for definitions tied to the virial radius of halos its total rate of change is a sum of the rate of change arising from the Hubble expansion and a term arising from the growth of the mass in the halo:
\begin{equation}
\dot{R}_{\rm s} = \frac{\partial R_{\rm s}}{\partial M_{\rm h}} 
\frac{dM_{\rm h}}{dt} + 
\frac{\partial R_{\rm s}}{\partial z} 
\frac{dz}{dt}.
\end{equation}
This derivation is completely general as long as $R_{s}=R_{s}(M_{\rm h},~z)$, and any redshift dependence of the density contrast  used to define $R_{s}$ is folded in the second term on the right-hand side. 
For $R_{\rm s}=R_{\rm vir}=R_{180b}$, straightforward algebra yields
\begin{equation}
\dot{R}_{180b} = R_{180b}
\left[
H(z) + \frac{\dot{M}_{180b}}{3 M_{180b}}
\right].
\end{equation}\\
We can use the fitting formula of \cite{2010MNRAS.406.2267F} (FMB10) for the average mass accretion rate of halos measured from a dark matter-only simulation to estimate the relative magnitude of the two terms in this expression. 
FMB10 found that the median mass accretion rate follows
\begin{align}
\label{FMB10 median}
\langle
\dot{M}_{180b}
\rangle_{\rm med}
\approx 25.3~{\rm M}_{\odot}{\rm~yr^{-1}}
& \left( \frac{M_{180b}}{10^{12}~{\rm M_{\odot}}} \right)^{1.1}
(1 + 1.65z) \\ \notag
& \times \sqrt{\Omega_{\rm m}(1+z)^{3} + \Omega_{\Lambda}},
\end{align}
where we use the $\approx$ sign to indicate the fact that FMB10 actually measured the growth of the FoF mass, and not directly the growth of $M_{180b}$, although the two closely follow each other, as we have emphasized above. 
Using this equation gives
\begin{equation}
\frac{\langle \dot{M}_{180b} \rangle_{\rm med}}{3 M_{180b} H(z)} \approx
0.12 \left( 
\frac{M_{180b}}{10^{12}~{\rm M_{\odot}}} 
\right)^{0.1}
(1 + 1.65z),
\end{equation}
implying that the Hubble expansion term dominates at low $z$ and for $M_{180b} \lesssim 10^{12}$ M$_{\odot}$, but not necessarily at higher mass and redshift.

To compare with previous results on halo mass growth, such as those of FMB10, the most relevant definition is a Lagrangian shell that follows the halo and grows with it. 
We therefore include both the Hubble expansion and the mass growth term in modeling the expanding virial shells. 
The mass growth term is calculated on a halo-by-halo basis, using the mass growth rate measured for each halo.  
Unless otherwise specified, we adopt $R_{\rm s}=R_{\rm vir}$ and $\Delta R_{\rm s}=0.2 \Delta R_{\rm vir}$ throughout, which we have found to be a good compromise between numerical noise and spatial resolution.

\section{RESULTS}
\label{numerical results}

\subsection{Accretion Rates}
\label{accretion rates}
We begin by quantifying the accretion rates of the dark matter, cold gas ($n_{\rm H}<0.13$ cm$^{-3}$, $T\leq2.5\times10^{5}$ K), hot gas ($n_{\rm H}<0.13$ cm$^{-3}$, $T>2.5\times10^{5}$ K), and ISM ($n_{\rm H}\geq0.13$ cm$^{-3}$) as a function of halo mass and redshift. 
For comparison, we also present results for the star formation rate within the halos in the simulations. 
Figure \ref{compare wind sims} shows the median \emph{net} accretion rates through the virial shell, for the simulations with different outflow parameterizations, at $z=0-5$.

As expected, the dark matter accretion rates are insensitive to the wind feedback, since they are not subject to hydrodynamical forces. 
The median net dark matter accretion rates from our hydrodynamic simulations are in excellent agreement with the fitting formula of FMB10 (eq. (\ref{FMB10 median})) based on the Millennium simulations, at all redshifts $z\geq1$. 
Differences between the median dark matter accretion rates of a factor $\approx1.5-2$ are apparent at $z=0$. 
However, our \emph{mean} dark matter accretion rates are in excellent agreement with the corresponding FMB10 fit at all redshifts, indicating (as we have verified) that it is the shape of the dark matter accretion rate distribution that is slightly different in our simulations. 
The dominant effect likely lies in the definition of the accretion rates. 
In this work, the accretion rates are instantaneous and directly evaluated from particle velocities in single snapshots. 
FMB10, on the other hand, evaluated the accretion rates by dividing the mass difference in subsequent snapshots by the time interval. 
At $z=0$, the Millennium snapshots are spaced by $\approx260$ Myr and so FMB10's accretion rates are averages over a comparable time scale. 
The effects of averaging are most pronounced at $z\sim0$, where accretion events are rarest, explaining why they are not as apparent at $z\geq1$. 
Additional modest differences are expected due to the different (first-year WMAP) cosmology assumed in the Millennium simulations. 

The baryonic rates, on the other hand, depend substantially on the wind prescription. 
In particular, the star formation and net cold gas accretion rates in low-mass halos are strongly suppressed as winds remove gas from the galaxies. 
In sufficiently massive halos, however, Figure \ref{compare wind sims} shows that
the accretion rates of the cold and hot gas are largely insensitive to the wind models
implemented in the simulations.  This independence arises in part because as halo escape velocity becomes sufficiently large, the outflowing material cannot reach the virial radius. 
As demonstrated in \cite{2008MNRAS.387..577O} and \cite{2010MNRAS.406.2325O}, hydrodynamic interactions with the gas surrounding the galaxy also help confining outflowing material in high-mass halos. 
These authors further showed that, as a result, massive halos preferentially confine their outflows even in models in which outflow velocity scales with the escape velocity of the halo, as expected in momentum-driven models \citep[e.g.,][]{2005ApJ...618..569M}. 

It is important to note that although the gas accretion rates through the virial radius appear insensitive to the details of the outflow prescription in massive halos, star formation feedback has been typically found to be insufficient in suppressing the galaxy baryon mass function to the observed level in these galaxies \citep[e.g.,][]{2010MNRAS.406.2325O}. 
Further feedback by active galactic nuclei (AGN), not included here, is probably responsible for suppressing the galaxy mass function in this mass regime \citep[e.g.,][]{2006MNRAS.365...11C, 2008MNRAS.391..481S, 2008ApJS..175..390H}.

Comparing the net and pure infall baryonic accretion rates is useful for assessing whether the net influx is affected by galactic winds solely because of the negative contribution from outflowing material, or whether infalling material can also be effectively kept out of halos via interactions with the outflows. 
If galactic winds simply eject material from the halos but do not significantly disrupt infalling baryonic structures, then the pure infall rates should be insensitive to the wind prescription, even if the net accretion rates depend strongly on it. 
This comparison is particularly instructive for the cold gas component, as it is predicted to provide most of the fuel for star formation \citep[e.g.,][]{2005MNRAS.363....2K, 2009MNRAS.395..160K}. 
As Figure \ref{Mdot cold in comparison} shows, the pure infall $\dot{M}_{\rm cold}^{\rm in}$ rates are indeed less sensitive to the wind prescription than the net rates (c.f., Fig. \ref{compare wind sims}), and relatively weakly dependent on it for the \verb|winds| and \verb|swinds| simulations, with $v_{\rm w}=342$ km s$^{-1}$. 
For our more extreme \verb|swinds| simulation with $v_{\rm w}=684$ km s$^{-1}$ and $\eta=2$, however, even the pure infall cold gas accretion rates are substantially affected at most redshifts, indicating that the outflows can in fact affect the infalling cold filaments. 
This is consistent with the inference of \cite{2010MNRAS.406.2325O} based on the suppression of the star formation rates by outflows (see also \S \ref{SFR Mdot cold section}), but is for the first time directly shown here. 

In Figure \ref{Mdot cold vs r}, we show how the net cold gas accretion rate varies with redshift and halo mass, for the no-wind simulations, but through shells of radii $R_{\rm s}=R_{\rm vir}$, $0.5R_{\rm vir}$, and $0.2R_{\rm vir}$ ($\Delta R_{s}=0.2R_{\rm vir},~0.2R_{\rm vir},~$ and $0.1R_{\rm vir}$). This illustrates how deep the cold gas penetrates into halos in different regimes. 
While at $z\geq2$ the cold gas accretion rate is significant down to $0.2R_{\rm vir}$ even in halos of mass $M_{\rm h}\gtrsim10^{11.5}$ M$_{\odot}$ that are dominated by hot gas (\S \ref{baryonic contents of halos}), the cold gas accretion rate at small radii is markedly suppressed in those halos at $z\leq1$. 
This confirms the persistence of cold streams into massive halos at high redshift \citep[][]{2005MNRAS.363....2K, 2009MNRAS.395..160K, 2008MNRAS.390.1326O}, arising from the short cooling times in the dense gas brought in along the dark matter filaments into rare halos \citep[][]{2006MNRAS.368....2D}. 
This phenomenon operates increasingly less efficiently at $z<2$ as the densities become lower, the cooling times correspondingly longer, and halos of a given mass become more abundant and tend to reside inside large dark matter filaments, rather than at their intersection \citep[][]{2003ASSL..281..185K}. 
In addition to shocking, the decreasing cold gas accretion rates with decreasing radius are caused by accretion by satellite galaxies on the way down to the center of the main halo \citep[e.g.,][]{2009MNRAS.395..160K, 2009MNRAS.399..650S}. 

In some regimes, particularly in high-mass halos, the hot gas accretion rates are found to be insensitive to the outflow prescription \citep[see also][]{2011MNRAS.414.2458V}. 
This indicates that outflows have a small impact on gas shocking at the virial radius. 
However, this does not hold universally and Figure \ref{compare wind sims} shows significant outflow dependences at lower masses, where winds can reach the virial radius with significant velocities. 

\begin{figure*}
\begin{center}
\includegraphics[width=1.0\textwidth]{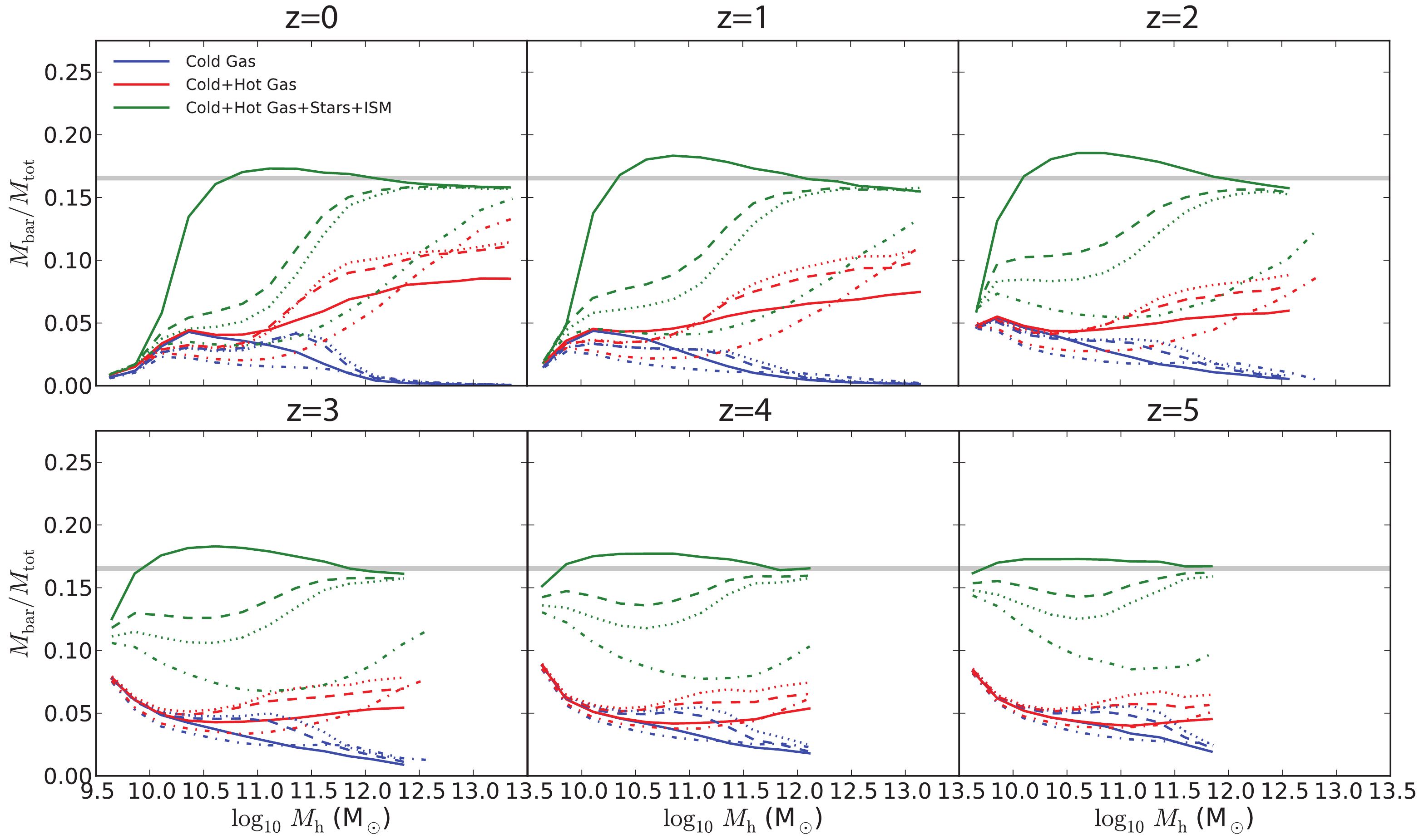}
\end{center}
\caption[]{Comparison of the median baryon mass fractions within halos, broken down by components, for the different wind prescriptions. We comment on the relative contributions of stellar and ISM material in \S \ref{baryonic contents of halos}. \emph{Solid curves}: no winds. \emph{Dashed:} constant-velocity winds with $v_{\rm w}=342$ km s$^{-1}$ and mass loading $\eta=1$ ({\tt winds}). 
\emph{Dotted:} constant-velocity winds with $v_{\rm w}=342$ km s$^{-1}$ and mass loading $\eta=2$ ({\tt swinds}).
\emph{Dash-dotted:} constant-velocity winds with $v_{\rm w}=684$ km s$^{-1}$ and mass loading $\eta=2$ ({\tt fwinds}).
The thick grey lines show the universal ratio $\Omega_{\rm b}/\Omega_{\rm m}$. 
}
\label{baryonic mass fractions} 
\end{figure*}

\subsection{Validity of Simple Scaling Relations}
\label{validity of scaling relations}
One motivation for quantifying the baryonic accretion rates onto halos is to provide 
more realistic ingredients for galaxy formation models, i.e. to extend the standard framework based on dark matter-only results that has been the backbone of most theoretical models so far. 
In this section, we assess the validity of several scaling relations that are often used to model gas accretion and star formation in halos based on dark matter-only models. 
Such scaling relations have been qualitatively motivated by the results of hydrodynamical simulations on the cold mode. 
Although limitations to such assumptions have already been clearly outlined \citep[for example, hot mode accretion is more important in some regimes; e.g.,][]{2005MNRAS.363....2K, 2009MNRAS.395..160K}, the level of their accuracy has not been systematically quantified over broad intervals of halo mass and redshift, and in the presence of outflows. 
We first review the motivations for such scaling relations, before exploring them in detail. 

Because the cold mode is characterized by infall that follows the dark matter filaments closely at large radii, it is tempting to assume cold gas enters halos at a rate following the dark matter with a universal baryonic mass factor, i.e. $\dot{M}_{\rm cold} + \dot{M}_{\rm ISM} \approx (\Omega_{\rm b} / \Omega_{\rm DM}) \dot{M}_{\rm DM}$. 
Furthermore, the cold mode is predicted to provide the main fuel for star formation.  
One might therefore assume that the star formation rates within halos are approximately equal to the rates at which they are supplied with cold gas, i.e., $SFR\approx \dot{M}_{\rm cold}+\dot{M}_{\rm ISM}$, as is the case globally in some regimes in the absence of feedback \citep[e.g.,][]{2005MNRAS.363....2K, 2009MNRAS.395..160K}. 
Together, these scalings lead to the common assumption that the star formation rates within halos is approximated by a universal baryonic factor of the rate at which they acquire dark matter: $SFR\approx (\Omega_{\rm b} / \Omega_{\rm DM}) \dot{M}_{\rm DM}$ \citep[e.g.,][]{2008ApJ...688..789G, 2009ApJ...703..785D, 2010ApJ...718.1001B}, with an optional efficiency factor. 
It is however clear that these simple scalings cannot apply universally, as the cold mode is suppressed in halos above the transition halo mass $M_{\rm t}\sim3\times10^{11}$ M$_{\odot}$ \citep[e.g.,][]{2003MNRAS.345..349B, 2005MNRAS.363....2K, 2009MNRAS.395..160K, 2008MNRAS.390.1326O}. 
As shown above, feedback can also strongly affect both the gaseous accretion and star formation rates.
It is therefore important to quantify the applicability of these simplified assumptions. 

\subsubsection{When Does $\dot{M}_{\rm cold} + \dot{M}_{\rm ISM} \approx (\Omega_{\rm b} / \Omega_{\rm DM}) \dot{M}_{\rm DM}$?}
\label{Mdot cold Mdot DM section}
Figure \ref{Mdot cold Mdot DM} compares the cold gas+ISM and dark matter accretion rates in our different simulations with and without galactic winds. 
While the gas accretion rates shown are dominated by the diffuse cold component, we have included the ISM contribution since most of it is also in cold form, and it can in principle be rapidly converted into stars as it falls into halos, either spontaneously or after merging with another galaxy. 
The $(\dot{M}_{\rm cold}+\dot{M}_{\rm ISM})/\dot{M}_{\rm DM}$ ratio is strongly reduced at halo masses $M_{\rm h}\gtrsim10^{11.5}$ M$_{\odot}$, as halos above this mass maintain stable virial shocks \citep[e.g.,][]{2003MNRAS.345..349B, 2005MNRAS.363....2K, 2009MNRAS.395..160K} and an increasing fraction of the infalling gas is therefore hot, as indicated by the red curves which include all the gas. In high-mass halos where the total gas to dark matter accretion rate ratio is significantly below universal, the accretion of stars make up most of the difference. 

At lower halo masses, $(\dot{M}_{\rm cold}+\dot{M}_{\rm ISM})/\dot{M}_{\rm DM}$ is further suppressed in simulations with outflows. 
As discussed in \S \ref{accretion rates} (c.f., Figs \ref{compare wind sims} and \ref{Mdot cold in comparison}), this is predominantly due to cold outflowing material contributing negatively to the virial shell flux, although hydrodynamical interactions between the inflowing and outflowing gas do suppress the infall rate of cold gas non-negligibly. 
At late times, Figure \ref{Mdot cold Mdot DM} shows that the median $(\dot{M}_{\rm cold}+\dot{M}_{\rm ISM})/\dot{M}_{\rm DM}$ is systematically below the universal value $\Omega_{\rm b}/\Omega_{\rm DM}$ even for low-mass halos that are not subject to strong virial shocking and in the absence of winds. 
Including the hot gas does not significantly change this ratio, and we have verified that further adding stars (i.e., including all the baryons) does not change the picture at the low-mass end, so that more than half of the halos in this regime accrete baryons at sub-universal rates. 
In the absence of outflows, the main mechanism keeping baryons out of low-mass halos is photoheating by the UV background \citep[e.g.,][]{1992MNRAS.256P..43E, 1996ApJ...465..608T, 2000ApJ...542..535G}. 
This effect is quantified in \S \ref{baryonic contents of halos} for our simulations, and is in fact sufficient to suppress the net accretion rate of baryons in halos with $M_{\rm h} \lesssim 10^{10.5}$ M$_{\odot}$ at $z\leq1$.

Interestingly, the $z=3-5$ panels of Figure \ref{Mdot cold Mdot DM} show that halos with $M_{\rm h}\lesssim 10^{11}$ M$_{\odot}$ can accrete cold gas at ``super-universal'' rates with respect to the dark matter, i.e. $(\dot{M}_{\rm cold}+\dot{M}_{\rm ISM})>\Omega_{\rm b}/\Omega_{\rm DM}$, for the simulation without outflows. 
The UV background is also likely responsible for this effect: the gas kept out of low-mass halos remains available for future accretion and can contribute to super-universal gas accretion rates in other halos. 
Net accretion of baryons rates $>(\Omega_{\rm b}/\Omega_{\rm DM})\dot{M}_{\rm DM}$ in some regimes is in fact required to explain the super-universal median baryon fractions over important intervals in halo mass and at all redshifts studied here, in the absence of outflows (\S \ref{baryonic contents of halos}). 
This effect, albeit small and only apparent in simulations without strong feedback, is also present in a number of other cosmological simulations using different codes \citep[][]{2002MNRAS.333..649S, keres_thesis} and is therefore believed to be real. 
The transition versus redshift from super- to sub-universal median baryonic accretion rates in the low-mass bin $M_{\rm h}\approx10^{10.5}$ M$_{\odot}$ can be explained by the redshift dependence of the UV background suppression scale discussed in \S \ref{baryonic contents of halos}. 
In models with outflows, the ejection of baryons from low-mass halos could also in principle permit subsequent super-universal accretion onto other halos, but this is not apparent in Figure \ref{Mdot cold Mdot DM}, indicating that it is subdominant to the first-order effect of winds ejecting gas from the halos.

\begin{figure*}
\begin{center} 
\includegraphics[width=1.0\textwidth]{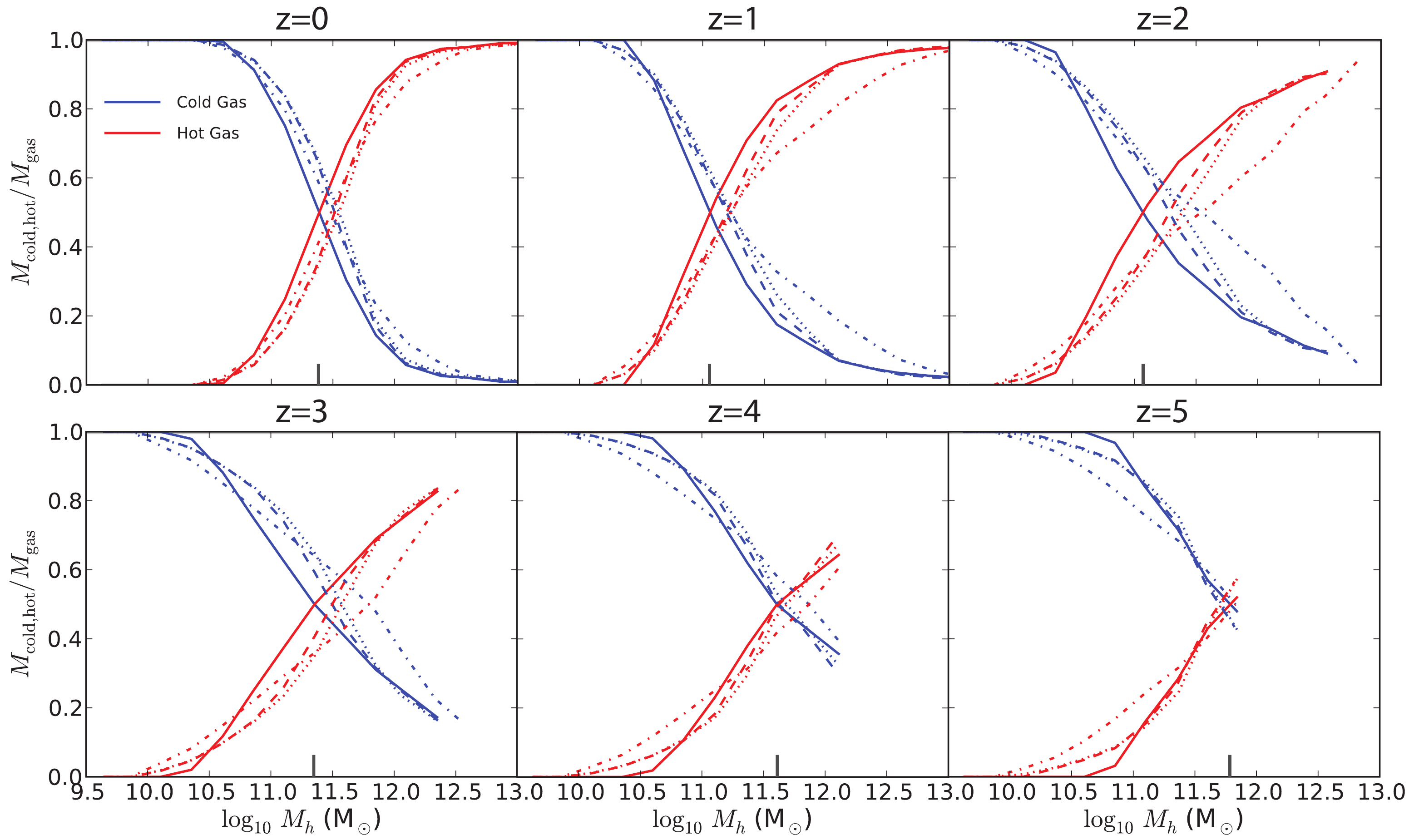}
\end{center}
\caption[]{Comparison of the cold and hot halo gas fractions, for the different wind prescriptions. 
\emph{Solid curves}: no winds. 
\emph{Dashed:} constant-velocity winds with $v_{\rm w}=342$ km s$^{-1}$ and mass loading $\eta=1$ ({\tt winds}). 
\emph{Dotted:} constant-velocity winds with $v_{\rm w}=342$ km s$^{-1}$ and mass loading $\eta=2$ ({\tt swinds}).
\emph{Dash-dotted:} constant-velocity winds with $v_{\rm w}=684$ km s$^{-1}$ and mass loading $\eta=2$ ({\tt fwinds}). 
Vertical markers indicate the halo mass at which the cold and hot gas mass fractions are equal for the no-wind case in each panel.}
\label{cold hot mass fractions} 
\end{figure*}

\subsubsection{When Does $SFR\approx(\dot{M}_{\rm cold}+\dot{M}_{\rm ISM})$?}
\label{SFR Mdot cold section}
Without strong feedback, star formation in galaxies is expected to follow the cosmological supply of gas. 
Observationally, galaxies otherwise exhaust their fuel on time scales $\sim1-2$ Gyr locally, with some evidence of even shorter gas-depletion time scales $\sim0.5$ Gyr at $z\sim2$ \citep[e.g.,][]{2010MNRAS.407.2091G}. 
Simulations adopting star formation laws motivated by observations in fact confirm that the star formation rates are limited by the accretion of intergalactic gas, and have further predicted that cold mode accretion is the main channel by which galaxies get their gas \citep[e.g.,][]{2002ApJ...571....1M, 2005MNRAS.363....2K, 2009MNRAS.395..160K}. 
This is illustrated in Figure \ref{SFR Mdot cold}, in which we plot the median ratio of the star formation rates within halos to the rates at which cold gas+ISM enter their virial radii, $SFR/(\dot{M}_{\rm cold}+\dot{M}_{\rm ISM})$. 
At redshifts $z\sim1-3$, the median ratio is close to unity at all resolved halo masses, i.e. for more than two decades in mass, in our simulation without outflows. 
This is the case even though a large fraction of the infalling gas in halos with $M_{\rm h}\gtrsim 10^{11.5}$ M$_{\odot}$ is hot, reflecting the minor contribution of hot mode accretion to star formation at this redshift \citep[][]{2009MNRAS.395..160K}. 

In our constant-velocity wind simulations, $SFR/(\dot{M}_{\rm cold} + \dot{M}_{\rm ISM})$ is heavily suppressed in low-mass halos at all redshifts, as outflows eject gas from galaxies before it can form stars. 
In sufficiently massive halos, the situation is reversed and the star formation rates in halos can exceed cold gas+ISM accretion rates by factors of a few (see also Fig. \ref{compare wind sims}). 
In all cases, we find that the $SFR/(\dot{M}_{\rm cold}+\dot{M}_{\rm ISM})$ ratio is enhanced by a factor $<(1+\eta)$, which is consistent with the additional star formation being fueled by the fall back of wind ejecta (in this regime, Fig. \ref{Mdot cold Mdot DM} shows that $\dot{M}_{\rm cold}+\dot{M}_{\rm ISM}$ is relatively unaffected by winds, so that variations in $SFR/(\dot{M}_{\rm cold}+\dot{M}_{\rm ISM})$ reflect variations in $SFR$). 
As expected, the high-velocity winds (\verb|fwinds|) contribute enhanced star formation only in halos significantly more massive than the slow winds (\verb|winds|, \verb|swinds|) do.
At late times ($z=0$), the ratio $SFR/(\dot{M}_{\rm cold}+\dot{M}_{\rm ISM})$ is suppressed much more heavily than the factor $1/(1+\eta)$ that would result if the only effect of outflows were to remove $\eta$ M$_{\odot}$ of galactic gas for reach 1 M$_{\odot}$ of stars formed. 
This again indicates that outflows are able to \emph{prevent} accretion of gas onto galaxies in this regime. 
Interestingly, the predicted $z=0$ star formation rates are lower than observationally-inferred ones in $\emph{all}$ of our wind models for $M_{\rm h}\lesssim10^{12}$ M$_{\odot}$. 
For example, at $z=0.1$, \cite{2009ApJ...696..620C} infer star formation rates $\sim3$ M$_{\odot}$ yr$^{-1}$ in $M_{\rm h}=10^{12}$ M$_{\odot}$ halos, and $\sim 0.2$ M$_{\odot}$ yr$^{-1}$ in $M_{\rm h}=10^{11}$ M$_{\odot}$ halos, based on abundance matching. 
At $z=0$, our simulations predict comparable star formation rates in $M_{\rm h}\approx10^{12}$ M$_{\odot}$ halos for all but the most extreme $\verb|fwinds|$ outflow implementation. 
For $M_{\rm h}\approx10^{11}$ M$_{\odot}$ halos, however, the $\verb|winds|$, $\verb|swinds|$, and $\verb|fwinds|$ simulations predict only $SFR\sim0.04,~0.02,~0.002$ M$_{\odot}$ yr$^{-1}$, respectively. 

The dominant factor explaining this discrepancy is likely the suppression of superwinds from real galaxies at $z\sim0$. 
Observations show that powerful superwinds are ubiquitous in galaxies with star formation rates per unit area $\Sigma_{\rm SFR}\geq0.1$ M$_{\odot}$ yr$^{-1}$ kpc$^{-2}$ \citep[e.g.,][]{2002ASPC..254..292H}, with outflows from other galaxies being significantly weaker \citep[e.g.,][]{2010AJ....140..445C}. 
Theoretically, this critical surface density can be derived from the condition that the massive star clusters in which most of the star formation of a galaxy occurs must be able to accelerate gas to velocities exceeding the escape velocity of the galaxy as a whole in order to produce a galactic-scale outflow \citep[][]{2011ApJ...735...66M}. 
Whereas this condition is satisfied in local starbursts and high-redshift Lyman break galaxies \citep[e.g.,][]{2006ApJ...647..128E}, it is usually not in present-day ordinary spirals, like the Milky Way. 
The high star formation rates permitted by the large cosmological infall rates at high redshift (Fig. \ref{compare wind sims}) ensure that superwinds operate effectively at early times. 
As the universe expands and becomes more tenuous, however, ordinary galaxies can no longer maintain star formation rates sufficient to drive significant outflows. 
In our simulations, this condition is not implemented and all galaxies drive outflows. 
As a result, the star formation rates at $z\sim0$ are artificially suppressed by feedback in the simulation with winds in some regimes. 
Note that star formation rates of the order of the observed ones are predicted in our simulation without winds at $M_{\rm h}\sim10^{11}-10^{12}$ M$_{\odot}$, indicating that cosmological infall is indeed sufficient to account for most of that star formation in the absence of feedback. 
Stellar mass loss, especially from AGB stars which can return up to $\sim50\%$ of the mass of a stellar population into the ISM over long time scales, is also likely to permit modestly higher star formation rates at late time \citep[e.g.,][]{2011ApJ...734...48L}, but is not modeled in our simulations aside for instantaneous gas recycling from Type II supernovae at the $10$\% level (\S \ref{galactic winds}). 
The high ISM metallicities that would result at $z=0$ if most star formation were fueled by stellar mass loss are however in tension with the measured ones, which argue for substantial recent accretion of metal-poor intergalactic gas \citep[e.g.,][]{2008A&ARv..15..189S}.

In high-mass halos, the situation is reversed and our simulations overpredict the observed star formation rates. 
This is the same problem alluded to in \S \ref{accretion rates} with respect to the baryon mass function, and is indicative of the need for additional feedback, most likely from AGN. 

Another interesting feature of Figure \ref{SFR Mdot cold} is that the ratio $SFR/(\dot{M}_{\rm cold}+\dot{M}_{\rm ISM})$ is significantly below unity even without feedback in halos of $M_{\rm h}\lesssim10^{11.5}$ M$_{\odot}$ at $z=5$, even though these halos are well into the cold mode-dominated regime.  
The likely explanation at this redshift, when the Universe just over 1 Gyr old, is that the Universe is so young that the recently formed low-mass galaxies have not had time to settle in a steady state in which the star formation rate follows the supply rate. 
In fact, the molecular gas depletion time scales, while yet to be directly measured at this redshift, are comparable to this time scale at $z\sim2$ \citep[e.g.,][]{2010MNRAS.407.2091G}. 
As Figure \ref{SFR Mdot cold} indicates, an approximate steady state between cold gas supply and star formation rate is reached by $z=3$ in our simulation without outflows. 

At $z=0$, the star formation rate is again slightly suppressed relative to the influx of cold gas through the virial radius. 
Multiple effects likely contribute to this. 
First, present-day low-mass galaxies are particularly prone to support extended gaseous discs, in which material accreted from the IGM can be temporarily stored, rather than rapidly converted into stars. 
At the low densities of outer discs, star formation is observed to become very inefficient, with depletion time scales up to $\sim10^{11}$ yr \citep[e.g.,][]{2010AJ....140.1194B}. 
The SFR therefore naturally deviates from the supply of cold gas when a substantial fraction of the infalling gas joins the outer parts.  
At larger masses, deviations between the cold gas+ISM accretion rate at the virial radius and the star formation rate within the halo are not surprising at $z=0$. 

To this discussion we must add that the star formation rates in our simulations depend to some extent on the sub-resolution star formation prescription (\S \ref{code details}). 
Furthermore, they are sensitive to the gas distribution in galaxies, such as the fraction of gas stored in low-density outer parts (Kere\v{s} et al., in prep.\nocite{keres_arepo}). 
The quantitative details of the star formation rate predictions should therefore be treated with appropriate caution. 
The general points that the star formation rates in galaxies are typically supply-limited and fueled by cold gas accretion, and the importance of the re-accretion of wind ejecta, are however robust. 

\subsubsection{When Does $SFR\approx (\Omega_{\rm b} / \Omega_{\rm DM}) \dot{M}_{\rm DM}$?}
\label{SFR Mdot DM section}
The $(\dot{M}_{\rm cold}+\dot{M}_{\rm ISM})-\dot{M}_{\rm DM}$ and $SFR-(\dot{M}_{\rm cold}+\dot{M}_{\rm ISM})$ relations discussed above combine to yield the $SFR-\dot{M}_{\rm DM}$ relation. 
This relation most directly quantifies the validity of the hypothesis that galaxies form stars at rates determined by the cosmological infall predicted in dark matter models. 
As Figure \ref{SFR Mdot DM} ($SFR/\dot{M}_{\rm DM}$ vs. halo mass and redshift) shows, the assumption that $SFR\approx (\Omega_{\rm b} / \Omega_{\rm DM}) \dot{M}_{\rm DM}$ is essentially never very accurate, even in simulations without galactic outflows. 
When outflows are included, the relation is substantially modified. 
In the constant-velocity models explored here, the two principal effects are to suppress the star formation rates in low-mass halos, but to boost them in more massive ones that can confine their outflows and later fuel their galaxies via re-accretion. 
The resulting $SFR-\dot{M}_{\rm DM}$ relation is however very sensitive to the uncertain outflow parameters, thus limiting the accuracy of simple estimates based on the dark matter infall. 

\subsection{Integrated Baryonic Halo Properties}
\label{baryonic contents of halos}
We now turn to quantifying some of the integrated baryonic properties of halos.
Since all the particles within the halos are included in these predictions, we relax the minimum number of dark matter particles to 64 (from 500 for the accretion rates) to cover a larger dynamic range. 
 
Figure \ref{baryonic mass fractions} shows how the baryonic mass in halos is divided between cold gas, hot gas, stars, and ISM, for our different simulations. 
In the no-wind case, the total baryon fraction (shown by the solid green curves) is close to the universal ratio $\Omega_{\rm b}/\Omega_{\rm m}$ above a redshift-dependent halo mass, corresponding to suppression by the UV background. 
The order of magnitude of this mass scale corresponds to the IGM temperature $T_{\rm IGM}\sim10^{4}$ K, multiplied by a factor of a few to account for adiabatic heating of the gas as it falls into halos, and is most directly related to the halo circular velocity \citep[e.g.,][]{1992MNRAS.256P..43E, 1996ApJ...465..608T}. 
\cite{1996ApJ...465..608T} showed that halos with circular velocity $v_{c} \lesssim 50$ km s$^{-1}$ can maintain pressures high enough to prevent the gas from collapsing with the dark matter. 
Defining $v_{c} \equiv (G M_{\rm h} / R_{\rm vir})^{1/2}$ and adopting $M_{\rm h}=M_{180b}$, $R_{\rm vir}=R_{180b}$ for consistency with the rest of this paper (\S \ref{halo identification}), 
\begin{equation}
M_{\rm h} = {\rm 10^{10}~M_{\odot}} \left( \frac{v_{c}}{\rm 50~km~s^{-1}} \right)^{3} \left( \frac{1+z}{4} \right)^{-3/2}. 
\end{equation}
This result shows how the UV background suppression mass scale should increase with time, as is indeed found to be the case in our simulations. 
We note that this crude analytic derivation is only meant to illustrate the principal effects seen in Figure \ref{baryonic mass fractions}. 
In reality, the details of this suppression are also sensitive to the spectrum of the UV background via its impact on the photoheating rate and on the cooling function \citep[e.g.,][]{2009ApJ...703.1416F}, and is subject to a time delay \citep[][]{2000ApJ...542..535G}.

In the no-wind case, the total baryon fraction actually exceeds the universal value $\Omega_{\rm b}/\Omega_{\rm m}$ by a small amount in some regimes. 
In \S \ref{Mdot cold Mdot DM section}, we found a similar effect, but for the accretion rates. 
It is now apparent in Figure \ref{baryonic mass fractions} that the super-universal baryon fractions occur in halos with mass just above the photoionization suppression mass scale, lending credence to the hypothesis that the gas kept out of low-mass halos by photoheating provides the material necessary for the super-universal baryonic accretion rates \citep[][]{keres_thesis}. 
In very massive halos, gas pressure in hot atmospheres may push out a small fraction of the baryons outside the virial radius and explain the slightly sub-universal baryon fractions \citep[][]{2009ASPC..419..347D}. 

Figure \ref{baryonic mass fractions} groups stars and ISM together to limit the number of curves shown. 
We conventionally refer to the combination as the ``galactic mass'', cautioning that the ISM boundary here only reflects the implementation of the multiphase sub-resolution model ($n_{\rm H}\geq0.13$ cm$^{-3}$; see \S \ref{definition of accretion rates}), and that gas stored in extended low-density discs may be missed by this measure. 
The relative contribution of the stellar and ISM components varies with both redshift and halo mass. 
At $z=0$, $>90$\% of the galactic mass is in stars for all halo masses. 
This fraction gradually diminishes with increasing redshift, and preferentially so in low-mass halos, whose galactic mass is increasingly dominated by ISM gas. 
At $z=3$, the stellar and ISM components divide the galactic mass equally at $M_{\rm h}\approx10^{10.3}$ M$_{\odot}$, with more massive halos having a greater mass in stars, and lower-mass halos having a greater mass in ISM. 
By $z=5$, the 50\% point is 0.5 dex higher, at $M_{\rm h}\approx10^{10.3}$ M$_{\odot}$. 
Although these numbers are for the no-wind simulation, the picture is qualitatively similar in the simulations with outflows. 

As expected from their effects on the accretion rates, the principal impact of outflows is to remove baryons from the halos, especially at low masses. 
At very early times, the impact of winds in low-mass halos is small in our simulations (e.g., at $M_{\rm h}<10^{10}$ M$_{\odot}$ in the $z=5$ panel of Fig. \ref{baryonic mass fractions}) because the young galaxies they host have not had time to form a lot of stars and to eject much material. 
By $z=0$, the suppression of the baryon fraction in low-mass halos is however dramatic.   
At sufficiently high masses, halos confine the winds and so the predicted total baryon fractions approach $\Omega_{\rm b}/\Omega_{\rm m}$, as in the no-wind case. 
In general, the baryon fraction is increasingly suppressed as either the mass loading factor $\eta$ or the wind velocity $v_{\rm w}$ is increased. 
At fixed halo mass and outflow parameters, winds are also most effective at suppressing the halo baryon fractions at low redshift. 
This occurs for the same basic reason explaining the enhanced efficiency of preventive feedback from photoionization. 
Namely, halos of a given mass are less dense at low redshift, which reduces both their escape velocity and the hydrodynamic drag they can provide. 

In terms of observations, the compilation of \cite{2010ApJ...719..119D} suggests that the actual halo baryon mass fractions are close to universal only in clusters with mass $M_{\rm h}\gtrsim 10^{14}$ M$_{\odot}$ ($v_{\rm c}\gtrsim600$ km s$^{-1}$) locally, and are increasingly suppressed below this mass. 
For halos below approximately this mass, which do not emit significantly in the X-rays, it is however generally not possible to directly measure the diffuse halo gas. The estimated baryon mass fractions are therefore principally based on galactic baryonic mass measurements, and part of the baryons may be ``missing'' only because of observational limitations. For most galaxy-sized halos, however, the detected baryons represent such a small fraction of the dark matter mass \citep[e.g.,][]{2005ApJ...632..859M, 2005ApJ...635...73H}, and the halo cooling times should be sufficiently short, that it is unlikely that all the missing baryons are simply undetected. 
In contrast, in our no-wind simulation, all halos with $M_{\rm h}\gtrsim 10^{11}$ M$_{\odot}$ have near-universal baryon fractions. 
Only our most extreme $\verb|fwinds|$ simulation comes close to suppressing the baryon fraction significantly in halos as massive as $M_{\rm h} \sim 10^{14}$ M$_{\odot}$ at $z=0$, again highlighting the need for very strong feedback. 
This simulation is however too effective at removing baryons from $L^{\star}$ galaxies (\S \ref{galactic winds}), indicating that realistic feedback must have a more complex mass dependence, and is likely to have multiple sources. 
\cite{2010MNRAS.406..822M}, for example, showed that the inclusion of AGN feedback, in addition to stellar feedback, can account for the suppression of the baryon mass fraction in groups with mass up to $M_{\rm h} \sim 10^{14}$ M$_{\odot}$. 

Figure \ref{cold hot mass fractions} considers only the diffuse halo gas (i.e., exclude ISM material) and compares the mass fractions of the cold and hot components, as a function of halo mass and redshift.  
In agreement with previous work, the diffuse gas mass in low-mass halos is completely cold-dominated, and it is completely hot-dominated in high-mass halos \citep[e.g.,][]{2005MNRAS.363....2K, 2009MNRAS.395..160K}. 
In the no-wind case, the halo mass at which the cold and hot halo gas mass fractions are equal, the transition mass $M_{\rm t}$, agrees well with the equivalent but lower-resolution runs of \cite{2009MNRAS.395..160K} at $z=0$ and $z=3$, with a value slightly below $M_{\rm h}=10^{11.5}$ M$_{\odot}$. 
The $z=4-5$ panels clearly show that $M_{\rm t}$ increases to higher values at higher redshifts, as the cooling times (particularly along the dense dark matter filaments feeding rare halos) are sufficiently short to permit the existence of cold streams even in halos that do develop virial shocks \citep[e.g.,][]{2005MNRAS.363....2K, 2009MNRAS.395..160K, 2006MNRAS.368....2D, 2008MNRAS.390.1326O}. 
A novel aspect of the present work is an exploration of several outflow parameterizations \citep[see also][]{2011MNRAS.414.2458V}. 
At $z=0$, the cold gas mass fraction in halos is seen to be nearly independent of the wind model, but some differences are seen at intermediate redshifts, particularly near $z=2$, where $M_{\rm t}$ in our most extreme outflow model (\verb|fwinds|) is higher by more than 0.5 dex relative to the no-wind case. 
The main effect of outflows in our simulations here seems to be to increase the cold gas mass fraction in halos by injecting them with cold gas from the ISM, and to consequently reduce the hot gas mass fraction.

Although the transition mass in terms of mass fractions is approximately constant, Figure \ref{compare wind sims} shows that the equivalent transition point for the accretion rates, i.e. the mass where $\dot{M}_{\rm cold}=\dot{M}_{\rm hot}$ at the virial radius, does evolve significantly with redshift \citep[see also][]{2008MNRAS.390.1326O}. 
For some applications, such as in semi-analytic models which require the fraction of the gas accreted by halos in different forms, this redshift-dependent transition point may be more relevant. 

\begin{figure}
\begin{center} 
\includegraphics[width=0.475\textwidth]{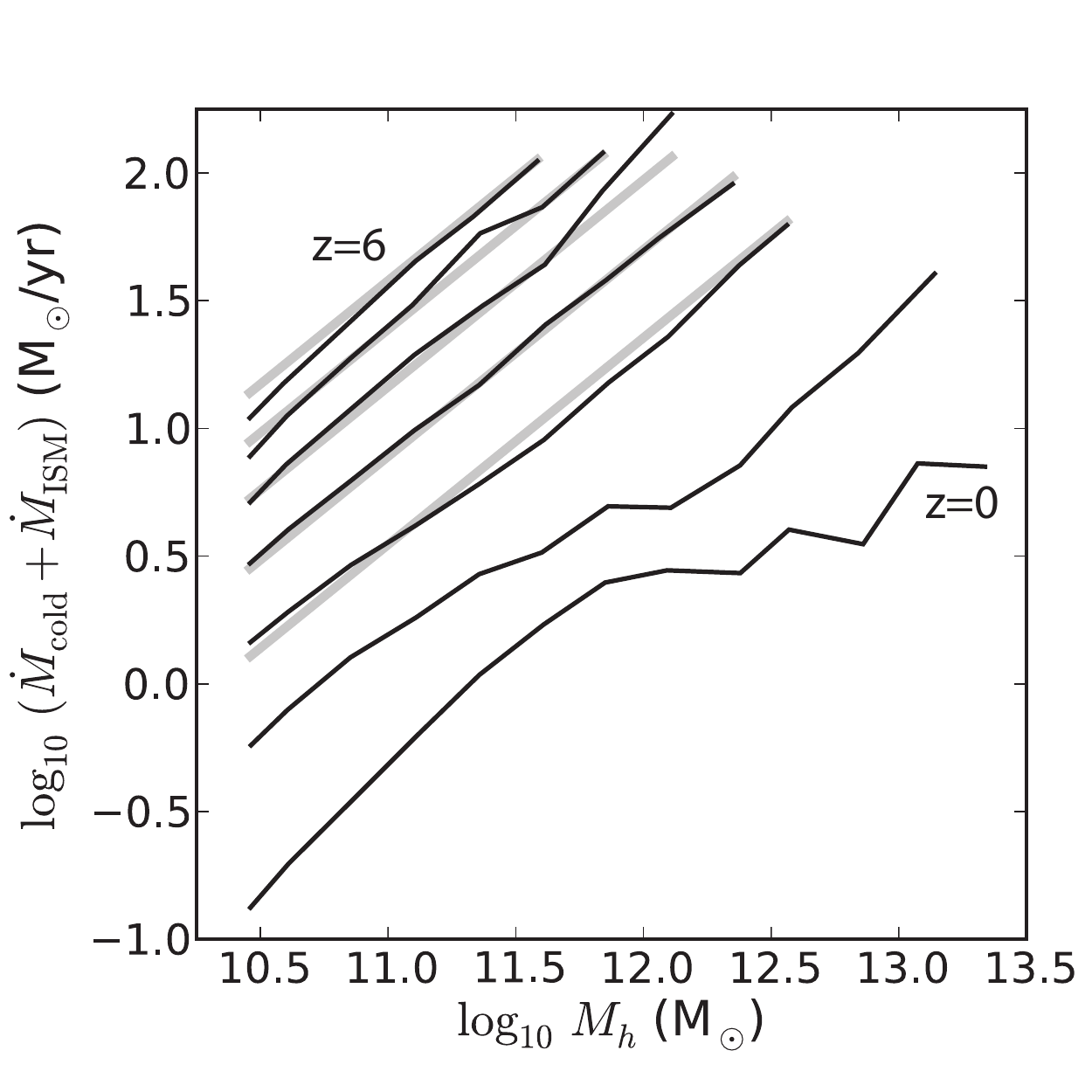}
\end{center}
\caption[]{Net cold gas+ISM accretion rates through the virial shell at $z=0,~1,~2,~3,~4,~5,$ and 6 (from the bottom up). The solid grey lines show a fit to the $z\geq 2$ data, where the star formation rates are most directly connected to the cold gas infall rates.
}
\label{Mdot cold fit} 
\end{figure}

\subsection{Fitting Formulae}
\label{fitting formulae}
While the results of this paper highlight the sensitivity of the baryonic accretion rates to the feedback model, the accretion rates in the no-wind case do not depend directly on model assumptions and may therefore be generally useful, provided that the effects of feedback are separately accounted for. 
Of these, the cold gas accretion rate is the most important as it provides the main fuel for star formation. 
As shown in \S \ref{accretion rates}, the infalling cold gas at the virial radius of massive halos is significantly suppressed during the infall at $z<2$. 
The star formation rates also differ significantly from the cold gas accretion rates at the virial radius at $z\sim0$. 
We therefore fit a simple model to the net $\dot{M}_{\rm cold}+\dot{M}_{\rm ISM}$ data versus halo mass and redshift, but limit ourselves to $z\geq2$, where it may be most robustly used to model star formation.

As it is useful to also consider the efficiency with which dark matter channels cold gas into halos, we begin by providing a fit to our median dark matter accretion rates (Fig. \ref{compare wind sims}). 
This fit accounts for the minor differences with the corresponding FMB10 formula arising from the definition of our accretion rates as being instantaneous and for the slightly different cosmology (\S \ref{accretion rates}). 
Whereas the Millennium simulations analyzed by FMB10 contained only dark matter, ours explicitly include baryons. 
Our dark matter accretion rates therefore, by definition, differ by a factor $(\Omega_{\rm m}-\Omega_{\rm b})/\Omega_{\rm m}$. 

We adopt the same functional form that was shown to yield an accurate fit by FMB10,
\begin{equation}
\label{FMB form}
\langle \dot{M} \rangle_{\rm med} = N (1 + \alpha z) \left( \frac{M_{\rm h}}{10^{12}~M_{\odot}} \right)^{\beta} \sqrt{\Omega_{\rm m}(1+z)^{3} + \Omega_{\Lambda}}.
\end{equation}
For the dark matter only, we find best-fit values $(N,~\alpha,~\beta)=(33.6$ M$_{\odot}$ yr$^{-1}$,~0.91,~1.06) to the data at $z=0,~1,~2,~3,~4,~5$ and 6 simultaneously. 
These fit the data in Figure \ref{compare wind sims} nearly perfectly. 
For the cold gas+ISM data, we fix the redshift evolution parameter $\alpha$ to the best-fit dark matter value, but allow for an additional overall redshift tilt factor $(1+z)^{\gamma}$. 
The resulting fit is as good as leaving $\alpha$ free, but yields a cleaner result for the efficiency below. 
The best-fit parameters for $\langle \dot{M}_{\rm cold} + \dot{M}_{\rm ISM} \rangle_{\rm med}$ ($z\geq2$ only) are then $(N,~\alpha,~\beta,~\gamma)=(1.84$ M$_{\odot}$ yr$^{-1}$,~0.91~0.81,~0.38).

These fits allow us to evaluate the efficiency 
\begin{equation}
\epsilon_{\rm cold} 
\equiv 
\frac{
\langle \dot{M}_{\rm cold} + \dot{M}_{\rm ISM} \rangle_{\rm med}
}{
\langle \dot{M}_{\rm DM} \rangle_{\rm med} [ \Omega_{\rm b}/(\Omega_{\rm m}-\Omega_{\rm b}) ]
},
\end{equation}
which would be unity if halos were supplied with cold gas+ISM with a universal baryon fraction of the dark matter accretion rate. 
Using the above fits, 
\begin{equation}
\label{epsilon cold fit}
\epsilon_{\rm cold}^{z\geq2} \approx 0.47 \left( \frac{1+z}{4} \right)^{0.38} \left( \frac{M_{\rm h}}{10^{12}~M_{\odot}} \right)^{-0.25}.
\end{equation}
Overall, the cold gas+ISM accretion rate has a shallower dependence on halo mass than the dark matter accretion rate. 
The owes principally to the greater ability of massive halos to main stable virial shocks, which heat a fraction of the gas (\S \ref{baryonic contents of halos}). 
These new fits to the cold gas+ISM accretion rates, which primarily drive star formation, provide a more realistic basis for galaxy formation models, and may have implications for studies that have assumed steeper mass slopes based on dark matter results \citep[e.g.,][]{2010ApJ...718.1001B}. 
Note that equation (\ref{epsilon cold fit}) would (unphysically) predict $\epsilon_{\rm cold} \to \infty$ as $z\to \infty$ or $M_{\rm h}\to 0$, so that the fits presented here should be understood to be valid only over the parameter space covered by the numerical data shown in Figure \ref{Mdot cold fit}, at the precision that the fit to $\langle \dot{M}_{\rm cold} + \dot{M}_{\rm ISM} \rangle_{\rm med}$ represent those data. 

Finally, it is instructive to compare Figure \ref{Mdot cold fit} (net $\dot{M}_{\rm cold}+\dot{M}_{\rm ISM}$) to Figure \ref{Mdot cold in comparison} (infalling $\dot{M}_{\rm cold}^{\rm in}$ component only). 
The similarity of the curves at corresponding redshifts indicates that the net cold gas accretion rates, in the no-wind simulation, are approximately equal to the infalling component only. 
Furthermore, the ISM component is a sub-dominant component of $\dot{M}_{\rm cold}+\dot{M}_{\rm ISM}$. 

\begin{figure}
\begin{center} 
\includegraphics[width=0.475\textwidth]{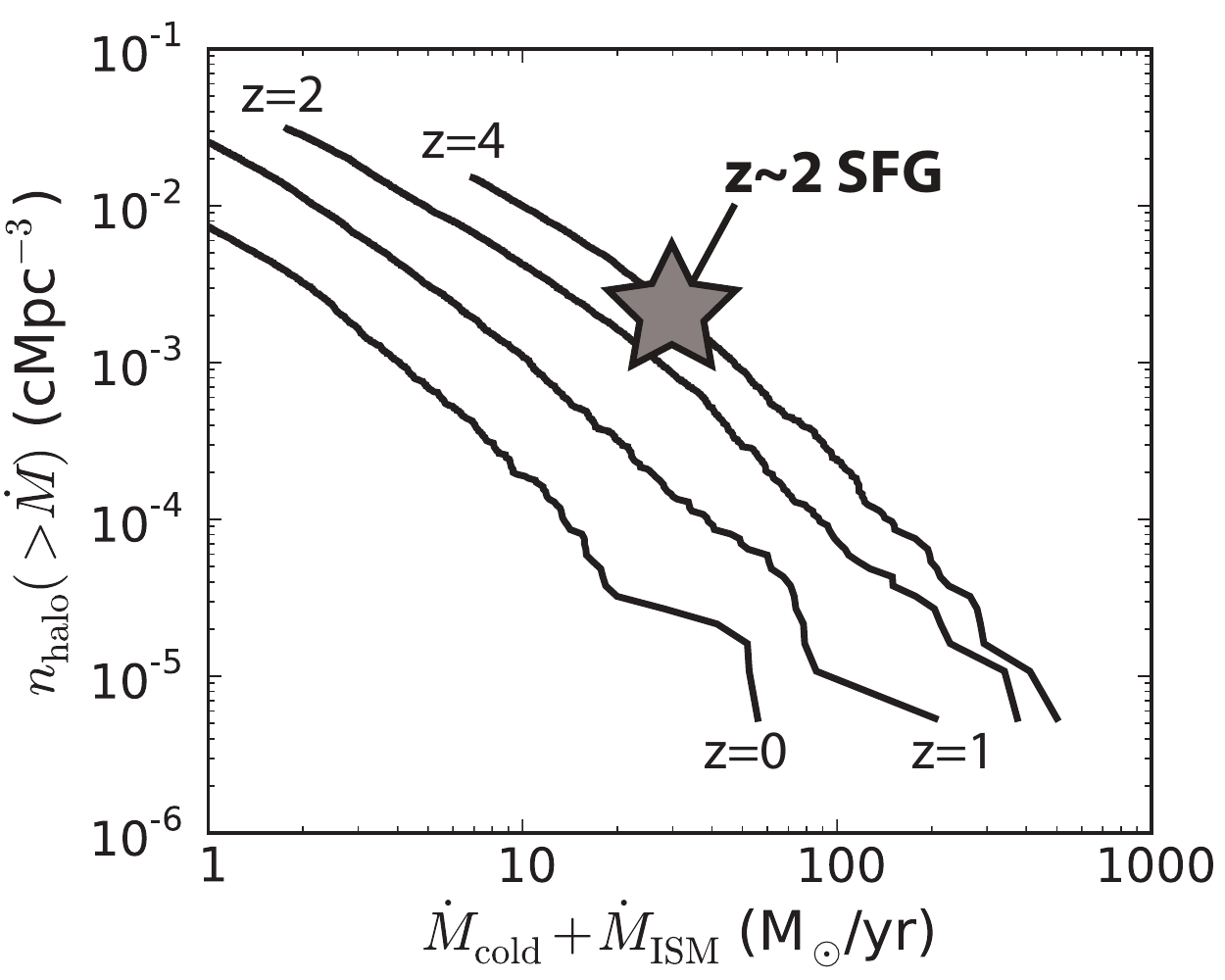}
\end{center}
\caption[]{Comoving number density of halos as a function of minimum net cold gas+ISM accretion rate through the virial shell in our no-wind simulation at $z=0,~1,~2,$ and 4 ($z=3$ omitted for clarity). The curves terminate at the highest accretion rates in the 40 $h^{-1}$ comoving Mpc box. For reference, the star shows the accretion rate equal to the median star formation rate in the H$\alpha$ sample of \cite{2006ApJ...647..128E} and the estimated the number density of BX galaxies based on clustering \citep[][]{2005ApJ...619..697A}.}
\label{nhalo} 
\end{figure}

\section{Discussion and Conclusion}
\label{discussion}
In this paper, we systematically quantified the rates at which dark matter halos accrete baryons and the resulting baryon mass fractions using a set of four high-resolution cosmological hydrodynamical simulations. 
Motivated by recent work indicating that star formation in galaxies is primarily driven by the supply of cold gas, we computed the gas accretion rates for the diffuse cold gas ($n_{\rm H}<0.13$ cm$^{-3}$, $T\leq2.5\times10^{5}$ K), hot gas ($n_{\rm H}<0.13$ cm$^{-3}$, $T>2.5\times10^{5}$ K), and ISM ($n_{\rm H}\geq0.13$ cm$^{-3}$) components separately, in addition to the dark matter. 
A main goal was to understand how existing results on the growth of the dark matter component of halos, which is accurately described by a set of analytic tools, including the Press-Schechter and excursion set formalisms \citep[][]{1974ApJ...187..425P, 1991ApJ...379..440B} and fitting formulae for their merger and accretion rates \citep[e.g.,][]{2002ApJ...568...52W, 2008MNRAS.383..615N, 2008MNRAS.386..577F, 2010MNRAS.406.2267F, 2009MNRAS.398.1858M, 2010ApJ...719..229G}, can be extended for baryons.  
Currently, dark matter-based analytic tools form the basis of many theoretical studies of galaxy formation, but it is usually necessary to make important assumptions to apply them to baryons. 

We investigated the baryonic accretion rates as a function of halo mass and redshift in a simulation without strong feedback, and for three otherwise identical simulations that implemented a phenomenological outflow model with different parameters (Fig. \ref{compare wind sims}). 
This allowed us to explore the sensitivity of our results to the uncertain feedback physics. 
By focusing on the infalling component only, we showed that winds can not only suppress the net accretion rates of cold gas by transporting cold gas outward, but also substantially affect the infalling cold material hydrodynamically, particularly in low-mass halos at low redshift (Fig. \ref{Mdot cold in comparison}). 
In halos sufficiently massive to confine their outflows owing to a combination of gravitational deceleration and hydrodynamic interactions, the accretion rates through the virial radius are however robust to the wind prescription. 
By evaluating the cold gas accretion rates through radial shells of radii ranging from $1R_{\rm vir}$ to $0.2R_{\rm vir}$ (Fig. \ref{Mdot cold vs r}), we explicitly showed how the accretion rate of cold gas diminishes as it penetrates into halos, owing to a combination of shocking and accretion onto satellite galaxies. 
Our results also confirm the persistence of cold streams into massive halos with hot atmospheres at $z\geq2$ \citep[][]{2005MNRAS.363....2K, 2009MNRAS.395..160K, 2006MNRAS.368....2D, 2008MNRAS.390.1326O} and their gradual disappearance a lower redshift. 

We directly tested the validity of a number of common simplifying assumptions, including whether the cold gas accretion rate traces the dark matter accretion rate with a universal baryon mass fraction (Fig. \ref{Mdot cold Mdot DM}), and whether the star formation rate within halos follows the cold gas accretion rate at the virial radius (Fig. \ref{SFR Mdot cold}). Ultimately, we quantified the accuracy with which the star formation rates in halos can be predicted from their dark matter accretion rates (Fig. \ref{SFR Mdot DM}).  
We find that in general such assumptions are not very accurate, highlighting the need for more accurate hydrodynamical results. 
In the absence of galactic outflows, the dark matter accretion rate is a predictor of the star formation rate at the factor of $\sim2$ level in halos below the transition mass $M_{\rm t}\sim10^{11.5}$ M$_{\odot}$ between the cold and hot modes. 
Above this transition mass, the dark matter accretion rate is an increasingly poor predictor of star formation, as the cooling times become longer than the free fall times, and gas entering the virial radius is no longer efficiently channeled into galaxies.  
This is especially so at lower redshift. 

When galactic outflows are included, the star formation rates are sensitive to the outflow parameters, especially in low-mass halos from which gas can be ejected altogether, where in some regimes star formation can be effectively stopped. 
In massive halos confining their outflows, the instantaneous star formation rates can on the other hand be boosted relative to the no-wind case by the re-accretion of gas onto galaxies. 

For our no-wind simulation, which is relatively free of \emph{ad hoc} assumptions, we have provided simple fitting formulae for the infall rates of cold gas through the virial radius, and for the corresponding efficiency with which dark matter channels cold gas into the halos (eq. (\ref{FMB form}-\ref{epsilon cold fit})). 
These will permit improvements of existing studies based on dark matter only.
 
An interesting consistency check is to compare the measured star formation rates at high redshift, where the high values are believed to be allowed by enhanced intergalactic accretion rates \citep[e.g.,][]{2009Natur.457..451D}, to the accretion rates of cold gas predicted by our simulations. 
We focus specifically on the $z\sim2$ star-forming galaxies that have been the subject of intense observational studies recently \citep[e.g.,][]{2004ApJ...604..534S, 2006ApJ...647..128E, 2007ApJ...670..156D}, and the targets of detailed integral field spectroscopic (IFS) analyses \citep[e.g.,][]{2006Natur.442..786G, 2009ApJ...706.1364F, 2009ApJ...697.2057L, 2009ApJ...699..421W}.
These IFS studies have revealed the $z\sim2$ disc galaxies to be very clumpy, which has been argued to be a direct consequence of high gas fractions maintained by rapid accretion \citep[e.g.,][]{2006ApJ...650..644E, 2009ApJ...703..785D, 2011ApJ...733..101G}. 
An exact comparison is complicated by the complex and sometimes uneven observational selection, and because the distribution of halo masses hosting the galaxies is not fully captured by the characteristic halo mass inferred from clustering. 
Furthermore, significant uncertainties exist in star formation rate estimates. 
For example, recent calorimetric SFR measurements by the \emph{Herschel Space Observatory} suggest that previous UV-based SFR estimates were biased high by a factor $\sim 2$ \citep[][]{2010A&A...518L..24N}. 
In general, satellite galaxies can also accrete material infalling through the virial radius of main halos, so that the halo star formation rates will be shared between the central galaxies and satellites. 
Keeping these caveats in mind, we therefore consider only the broad picture here. 

The median star formation rate in the IFS sample of \cite{2009ApJ...706.1364F} is $SFR\approx70$ M$_{\odot}$ yr$^{-1}$. 
The clustering of galaxies in a similarly selected parent sample indicates a characteristic halo mass $M_{\rm h}=10^{11.8}-10^{12.2}$ M$_{\odot}$ \citep[][]{2005ApJ...619..697A}. 
On the other hand, our fit to the $z\geq2$ simulation data (\S \ref{fitting formulae}) implies that the cold gas accretion rate through the virial radius, including ISM material, is $\approx23$ M$_{\odot}$ yr$^{-1}$ for a $M_{\rm h}=10^{12}$ M$_{\odot}$ halo at $z=2$ in the absence of outflows. 
While this is short by a factor $\sim3$ to explain the median $SFR\approx70$ M$_{\odot}$ yr$^{-1}$, the selection criteria of the IFS sample favor above-average systems permitting spatially-resolved spectroscopy \citep[][]{2009ApJ...706.1364F}. 
The 90$^{\rm th}$ percentile $\dot{M}_{\rm cold}+\dot{M}_{\rm ISM}$ accretion rate for this halo mass and redshift is in fact a factor $\approx3$ higher than the median. 

An alternative comparison is with the H$\alpha$ sample of \cite{2006ApJ...647..128E}, which was drawn from the larger rest UV-selected sample described by \cite{2004ApJ...604..534S},\footnote{UV selection techniques are designed to discover star-forming galaxies and therefore miss passive galaxies \citep[e.g.,][]{2006ApJ...638L..59V}. This population may account for as much as half of the massive galaxies at $z\sim2$, but their very low star formation rates \citep[e.g.,][]{2006ApJ...649L..71K} suggest that processes not included in our simulations (such as feedback during galaxy mergers) terminated star formation in them. Provided that, at a given halo mass, this putative quenching process is stochastic in the sense that it does not depend significantly on the cold gas accretion rate, it is fair to compare our simulation results with the sub-sample of star formers.} and which has a comparable characteristic halo mass $M_{\rm h}\sim10^{12}$ M$_{\odot}$. 
The median $SFR\approx30$ M$_{\odot}$ yr$^{-1}$ of this sample is in good agreement with the predicted supply rate. 
A potential worry is that the outflows inferred in these systems, with mass loading factors up to a few times their star formation rates \citep[e.g.,][]{2006ApJ...644..813E, 2008ApJ...674..151E}, might require commensurately higher accretion rates to sustain the star formation. 
However, as Figures \ref{compare wind sims} and \ref{SFR Mdot DM} show, the halos are sufficiently massive that, for realistic parameters, outflows can actually boost the star formation rates via gas re-accretion. 
Figure \ref{nhalo} also shows that the number density of BX galaxies \citep[][]{2005ApJ...619..697A}, from which Erb et al.'s sample is drawn, agrees well with the number density of halos with cold gas+ISM accretion rates sufficient to fuel their star formation in our no-wind simulation at $z=2$. 
Gas recycling from supernovae and AGB stars can also sustain star formation rates exceeding the external supply rate by $\sim50$\%, and hot mode accretion can provide some extra fuel \citep[e.g.,][]{2005MNRAS.363....2K, 2009MNRAS.395..160K}. 

While these considerations show that the high $SFR\sim10-100$ M$_{\odot}$ yr$^{-1}$ of $z\sim2$ galaxies are, broadly, well explained by being primarily determined by cold mode accretion, the quantitative details should be interpreted with caution. 
In particular, typical samples \citep[e.g.,][]{2006ApJ...647..128E, 2007ApJ...670..156D} cover up to two orders of magnitude in stellar mass, so that comparisons with characteristic halo masses are too simplistic to draw firm conclusions.  
The more intense star-formers with $SFR\gtrsim100$ M$_{\odot}$ yr$^{-1}$ (including sub-millimeter galaxies; e.g., Hayward et al. 2011\nocite{2011arXiv1101.0002H}) are most likely possible owing to a combination of variance in the accretion rates and other fueling channels, such as gas recycling, hot mode accretion, and mergers.\footnote{During major mergers, gas stored in the galaxies can be efficiently consumed in a nuclear starburst, resulting in an instantaneous star formation rate that can greatly exceed the cosmological infall rate \citep[e.g.,][]{1996ApJ...464..641M}.} 
A more detailed comparison of these with the results of simulations would be an interesting topic for future work, but is challenging owing to observational uncertainties. 

At $z=0$, we have shown that cosmological accretion can also account for the star formation rates inferred in $M_{\rm h}\sim10^{11}-10^{12}$ M$_{\odot}$ halos \citep[e.g.,][]{2009ApJ...696..620C}. 
Our simulations with outflows however all under-predict the star formations rates at $M_{\rm h}\sim10^{11}$ M$_{\odot}$, by factors ranging from 5 to 100. 
Most likely, this is because whereas all star-forming galaxies at high redshift drive powerful superwinds, this is not true at $z\sim0$. 
Instead, it appears that only galaxies with star formation rate surface density $\Sigma_{\rm SFR}\geq0.1$ M$_{\odot}$ yr$^{-1}$ kpc$^{-2}$ have superwinds \citep[e.g.,][]{2002ASPC..254..292H}. 
This criterion is usually satisfied at high redshift, where the star formation rates are high, but locally only in starbursts. 
Such a critical star formation rate surface density may be related to the ability of feedback processes to eject gas from the host galaxy \citep[e.g.,][]{2011ApJ...735...66M} and is not modeled in our phenomenological outflow implementation. 
It is also missing in parametric momentum-driven wind scalings that do not self-consistently model the wind generation \citep[e.g.,][]{2008MNRAS.387..577O, 2010MNRAS.406.2325O}. 
This suggests that more realistic wind models will be necessary to reproduce the observed local galaxy population. 

Without winds, the total baryon fraction in halos is close to universal in all halos above the UV background photoheating suppression scale, which our simulations resolve at all redshifts $z\lesssim5$ (Fig. \ref{baryonic mass fractions}). 
The UV background suppression mass scale is seen to shift to progressively higher masses with time (from $\sim10^{10}$ M$_{\odot}$ at $z=3$ to $\sim10^{10.5}$ M$_{\odot}$ at $z=0$), reflecting the fact that the most relevant halo property is the circular velocity, which increases with redshift at fixed halo mass. 
Such universal baryon fractions are in contrast with observations, which suggest that only galaxy clusters with $M_{\rm h}\gtrsim10^{14}$ M$_{\odot}$ may have near-universal baryon fractions at $z\sim0$ \citep[e.g.,][but see the discussion of the uncertainties on this mass scale in \S \ref{baryonic contents of halos}]{2010ApJ...719..119D}. 
Outflows are effective at suppressing the total baryon fractions up to a halo mass depending principally on the outflow velocity. 
Only our most extreme $\verb|fwinds|$ outflow prescription, with $v_{\rm w}=684$ km s$^{-1}$, suppresses the baryon fraction in halos close to the necessary scale. 
This again underscores the need for strong feedback to match the observations, likely from AGN in addition to stars. 

While the transition halo mass $M_{\rm t}\sim10^{11.5}$ M$_{\odot}$ at which halos host an equal mass fraction of cold and hot gas is approximately constant with redshift and insensitive to the outflow prescription (Fig. \ref{cold hot mass fractions}), the corresponding transition point in the accretion rates at the virial radius does depend significantly on redshift. 

Finally, these results should be useful to guide the implementation of cold accretion in semi-analytic models of galaxy formation \citep[e.g.,][]{2006MNRAS.370.1651C, 2009ApJ...700L..21K, 2011MNRAS.tmp.1035L, 2011MNRAS.410.2653B}, in particular in understanding how galactic outflows modify the standard results from hydrodynamical simulations. 
So far, the wind models we have explored have been phenomenological and the parameters were chosen to cover a representative range of possibilities. 
In future work, we will study the properties of self-consistently generated outflows \citep[][]{2011arXiv1101.4940H}, and the details of their interactions with infalling material.

\section*{Acknowledgments} 
We thank Volker Springel for providing the enhanced version of his GADGET code and Eliot Quataert for useful discussions.
CAFG is supported by a fellowship from the Miller Institute for Basic Research in Science, and received further support from the Harvard Merit Fellowship and FQRNT during the course of this work. 
DK is supported by NASA through Hubble Fellowship grant HST-HF-51276.01-A awarded by the Space Telescope Science Institute, which is operated by the Association of Universities for Research in Astronomy, Inc., under contract NAS 5-26555. CPM is supported by NASA through grant 10-ATP10-0187 and grant HST-AR-12140.01-A from the Space Telescope Science Institute, which is operated by the Association of Universities for Research in Astronomy, Inc., under contract NAS 5-26555. CPM is also supported by the Miller Institute for Basic
Research in Science, University of California Berkeley.
The computations presented in this paper were performed on the Odyssey cluster at Harvard University.

\bibliography{references} 

\begin{thebibliography}{}

\bibitem[\protect\citeauthoryear{{Adelberger}, {Steidel}, {Pettini}, {Shapley},
  {Reddy} \& {Erb}}{{Adelberger} et~al.}{2005}]{2005ApJ...619..697A}
{Adelberger} K.~L.,  {Steidel} C.~C.,  {Pettini} M.,  {Shapley} A.~E.,  {Reddy}
  N.~A.,    {Erb} D.~K.,  2005, \apj, 619, 697

\bibitem[\protect\citeauthoryear{{Barnes} \& {Hut}}{{Barnes} \&
  {Hut}}{1986}]{1986Natur.324..446B}
{Barnes} J.,  {Hut} P.,  1986, \nat, 324, 446

\bibitem[\protect\citeauthoryear{{Bauermeister}, {Blitz} \&
  {Ma}}{{Bauermeister} et~al.}{2010}]{2010ApJ...717..323B}
{Bauermeister} A.,  {Blitz} L.,    {Ma} C.,  2010, \apj, 717, 323

\bibitem[\protect\citeauthoryear{{Bell}, {McIntosh}, {Katz} \&
  {Weinberg}}{{Bell} et~al.}{2003}]{2003ApJS..149..289B}
{Bell} E.~F.,  {McIntosh} D.~H.,  {Katz} N.,    {Weinberg} M.~D.,  2003, \apjs,
  149, 289

\bibitem[\protect\citeauthoryear{{Benson}}{{Benson}}{2010}]{2010arXiv1008.1786B}
{Benson} A.~J.,  2010, ArXiv e-prints

\bibitem[\protect\citeauthoryear{{Benson} \& {Bower}}{{Benson} \&
  {Bower}}{2011}]{2011MNRAS.410.2653B}
{Benson} A.~J.,  {Bower} R.,  2011, \mnras, 410, 2653

\bibitem[\protect\citeauthoryear{{Bigiel}, {Leroy}, {Walter}, {Blitz},
  {Brinks}, {de Blok} \& {Madore}}{{Bigiel} et~al.}{2010}]{2010AJ....140.1194B}
{Bigiel} F.,  {Leroy} A.,  {Walter} F.,  {Blitz} L.,  {Brinks} E.,  {de Blok}
  W.~J.~G.,    {Madore} B.,  2010, \aj, 140, 1194

\bibitem[\protect\citeauthoryear{{Birnboim} \& {Dekel}}{{Birnboim} \&
  {Dekel}}{2003}]{2003MNRAS.345..349B}
{Birnboim} Y.,  {Dekel} A.,  2003, \mnras, 345, 349

\bibitem[\protect\citeauthoryear{{Bond}, {Cole}, {Efstathiou} \&
  {Kaiser}}{{Bond} et~al.}{1991}]{1991ApJ...379..440B}
{Bond} J.~R.,  {Cole} S.,  {Efstathiou} G.,    {Kaiser} N.,  1991, \apj, 379,
  440

\bibitem[\protect\citeauthoryear{{Bouch{\'e}}, {Dekel}, {Genzel}, {Genel},
  {Cresci}, {F{\"o}rster Schreiber}, {Shapiro}, {Davies} \&
  {Tacconi}}{{Bouch{\'e}} et~al.}{2010}]{2010ApJ...718.1001B}
{Bouch{\'e}} N.,  {Dekel} A.,  {Genzel} R.,  {Genel} S.,  {Cresci} G.,
  {F{\"o}rster Schreiber} N.~M.,  {Shapiro} K.~L.,  {Davies} R.~I.,
  {Tacconi} L.,  2010, \apj, 718, 1001

\bibitem[\protect\citeauthoryear{{Bower}, {Benson}, {Malbon}, {Helly}, {Frenk},
  {Baugh}, {Cole} \& {Lacey}}{{Bower} et~al.}{2006}]{2006MNRAS.370..645B}
{Bower} R.~G.,  {Benson} A.~J.,  {Malbon} R.,  {Helly} J.~C.,  {Frenk} C.~S.,
  {Baugh} C.~M.,  {Cole} S.,    {Lacey} C.~G.,  2006, \mnras, 370, 645

\bibitem[\protect\citeauthoryear{{Boylan-Kolchin}, {Springel}, {White},
  {Jenkins} \& {Lemson}}{{Boylan-Kolchin} et~al.}{2009}]{2009MNRAS.398.1150B}
{Boylan-Kolchin} M.,  {Springel} V.,  {White} S.~D.~M.,  {Jenkins} A.,
  {Lemson} G.,  2009, \mnras, 398, 1150

\bibitem[\protect\citeauthoryear{{Brooks}, {Governato}, {Quinn}, {Brook} \&
  {Wadsley}}{{Brooks} et~al.}{2009}]{2009ApJ...694..396B}
{Brooks} A.~M.,  {Governato} F.,  {Quinn} T.,  {Brook} C.~B.,    {Wadsley} J.,
  2009, \apj, 694, 396

\bibitem[\protect\citeauthoryear{{Cattaneo}, {Dekel}, {Devriendt}, {Guiderdoni}
  \& {Blaizot}}{{Cattaneo} et~al.}{2006}]{2006MNRAS.370.1651C}
{Cattaneo} A.,  {Dekel} A.,  {Devriendt} J.,  {Guiderdoni} B.,    {Blaizot} J.,
   2006, \mnras, 370, 1651

\bibitem[\protect\citeauthoryear{{Cattaneo}, {Mamon}, {Warnick} \&
  {Knebe}}{{Cattaneo} et~al.}{2010}]{2010arXiv1002.3257C}
{Cattaneo} A.,  {Mamon} G.~A.,  {Warnick} K.,    {Knebe} A.,  2010, ArXiv
  e-prints

\bibitem[\protect\citeauthoryear{{Chen}, {Tremonti}, {Heckman}, {Kauffmann},
  {Weiner}, {Brinchmann} \& {Wang}}{{Chen} et~al.}{2010}]{2010AJ....140..445C}
{Chen} Y.,  {Tremonti} C.~A.,  {Heckman} T.~M.,  {Kauffmann} G.,  {Weiner}
  B.~J.,  {Brinchmann} J.,    {Wang} J.,  2010, \aj, 140, 445

\bibitem[\protect\citeauthoryear{{Conroy} \& {Wechsler}}{{Conroy} \&
  {Wechsler}}{2009}]{2009ApJ...696..620C}
{Conroy} C.,  {Wechsler} R.~H.,  2009, \apj, 696, 620

\bibitem[\protect\citeauthoryear{{Croton}, {Springel}, {White}, {De Lucia},
  {Frenk}, {Gao}, {Jenkins}, {Kauffmann}, {Navarro} \& {Yoshida}}{{Croton}
  et~al.}{2006}]{2006MNRAS.365...11C}
{Croton} D.~J.,  {Springel} V.,  {White} S.~D.~M.,  {De Lucia} G.,  {Frenk}
  C.~S.,  {Gao} L.,  {Jenkins} A.,  {Kauffmann} G.,  {Navarro} J.~F.,
  {Yoshida} N.,  2006, \mnras, 365, 11

\bibitem[\protect\citeauthoryear{{Daddi}, {Dickinson}, {Morrison}, {Chary},
  {Cimatti}, {Elbaz}, {Frayer}, {Renzini}, {Pope}, {Alexander}, {Bauer},
  {Giavalisco}, {Huynh}, {Kurk} \& {Mignoli}}{{Daddi}
  et~al.}{2007}]{2007ApJ...670..156D}
{Daddi} E.,  {Dickinson} M.,  {Morrison} G.,  {Chary} R.,  {Cimatti} A.,
  {Elbaz} D.,  {Frayer} D.,  {Renzini} A.,  {Pope} A.,  {Alexander} D.~M.,
  {Bauer} F.~E.,  {Giavalisco} M.,  {Huynh} M.,  {Kurk} J.,    {Mignoli} M.,
  2007, \apj, 670, 156

\bibitem[\protect\citeauthoryear{{Dai}, {Bregman}, {Kochanek} \& {Rasia}}{{Dai}
  et~al.}{2010}]{2010ApJ...719..119D}
{Dai} X.,  {Bregman} J.~N.,  {Kochanek} C.~S.,    {Rasia} E.,  2010, \apj, 719,
  119

\bibitem[\protect\citeauthoryear{{Dav{\'e}}}{{Dav{\'e}}}{2009}]{2009ASPC..419..347D}
{Dav{\'e}} R.,  2009, in {S.~Jogee, I.~Marinova, L.~Hao, \& G.~A.~Blanc} ed.,
  Astronomical Society of the Pacific Conference Series Vol.~419 of
  Astronomical Society of the Pacific Conference Series, {Missing Halo Baryons
  and Galactic Outflows}.
p.~347

\bibitem[\protect\citeauthoryear{{Davis}, {Efstathiou}, {Frenk} \&
  {White}}{{Davis} et~al.}{1985}]{1985ApJ...292..371D}
{Davis} M.,  {Efstathiou} G.,  {Frenk} C.~S.,    {White} S.~D.~M.,  1985, \apj,
  292, 371

\bibitem[\protect\citeauthoryear{{Dekel} \& {Birnboim}}{{Dekel} \&
  {Birnboim}}{2006}]{2006MNRAS.368....2D}
{Dekel} A.,  {Birnboim} Y.,  2006, \mnras, 368, 2

\bibitem[\protect\citeauthoryear{{Dekel}, {Birnboim}, {Engel}, {Freundlich},
  {Goerdt}, {Mumcuoglu}, {Neistein}, {Pichon}, {Teyssier} \& {Zinger}}{{Dekel}
  et~al.}{2009}]{2009Natur.457..451D}
{Dekel} A.,  {Birnboim} Y.,  {Engel} G.,  {Freundlich} J.,  {Goerdt} T.,
  {Mumcuoglu} M.,  {Neistein} E.,  {Pichon} C.,  {Teyssier} R.,    {Zinger} E.,
   2009, \nat, 457, 451

\bibitem[\protect\citeauthoryear{{Dekel}, {Sari} \& {Ceverino}}{{Dekel}
  et~al.}{2009}]{2009ApJ...703..785D}
{Dekel} A.,  {Sari} R.,    {Ceverino} D.,  2009, \apj, 703, 785

\bibitem[\protect\citeauthoryear{{Dijkstra} \& {Loeb}}{{Dijkstra} \&
  {Loeb}}{2009}]{2009MNRAS.400.1109D}
{Dijkstra} M.,  {Loeb} A.,  2009, \mnras, 400, 1109

\bibitem[\protect\citeauthoryear{{Dutton}, {van den Bosch} \& {Dekel}}{{Dutton}
  et~al.}{2010}]{2010MNRAS.405.1690D}
{Dutton} A.~A.,  {van den Bosch} F.~C.,    {Dekel} A.,  2010, \mnras, 405, 1690

\bibitem[\protect\citeauthoryear{{Efstathiou}}{{Efstathiou}}{1992}]{1992MNRAS.256P..43E}
{Efstathiou} G.,  1992, \mnras, 256, 43P

\bibitem[\protect\citeauthoryear{{Efstathiou}}{{Efstathiou}}{2000}]{2000MNRAS.317..697E}
{Efstathiou} G.,  2000, \mnras, 317, 697

\bibitem[\protect\citeauthoryear{{Eisenstein} \& {Hu}}{{Eisenstein} \&
  {Hu}}{1999}]{1999ApJ...511....5E}
{Eisenstein} D.~J.,  {Hu} W.,  1999, \apj, 511, 5

\bibitem[\protect\citeauthoryear{{Elmegreen} \& {Elmegreen}}{{Elmegreen} \&
  {Elmegreen}}{2006}]{2006ApJ...650..644E}
{Elmegreen} B.~G.,  {Elmegreen} D.~M.,  2006, \apj, 650, 644

\bibitem[\protect\citeauthoryear{{Erb}}{{Erb}}{2008}]{2008ApJ...674..151E}
{Erb} D.~K.,  2008, \apj, 674, 151

\bibitem[\protect\citeauthoryear{{Erb}, {Shapley}, {Pettini}, {Steidel},
  {Reddy} \& {Adelberger}}{{Erb} et~al.}{2006}]{2006ApJ...644..813E}
{Erb} D.~K.,  {Shapley} A.~E.,  {Pettini} M.,  {Steidel} C.~C.,  {Reddy} N.~A.,
     {Adelberger} K.~L.,  2006, \apj, 644, 813

\bibitem[\protect\citeauthoryear{{Erb}, {Steidel}, {Shapley}, {Pettini},
  {Reddy} \& {Adelberger}}{{Erb} et~al.}{2006}]{2006ApJ...647..128E}
{Erb} D.~K.,  {Steidel} C.~C.,  {Shapley} A.~E.,  {Pettini} M.,  {Reddy} N.~A.,
     {Adelberger} K.~L.,  2006, \apj, 647, 128

\bibitem[\protect\citeauthoryear{{Fakhouri} \& {Ma}}{{Fakhouri} \&
  {Ma}}{2008}]{2008MNRAS.386..577F}
{Fakhouri} O.,  {Ma} C.,  2008, \mnras, 386, 577

\bibitem[\protect\citeauthoryear{{Fakhouri} \& {Ma}}{{Fakhouri} \&
  {Ma}}{2009}]{2009MNRAS.394.1825F}
{Fakhouri} O.,  {Ma} C.,  2009, \mnras, 394, 1825

\bibitem[\protect\citeauthoryear{{Fakhouri} \& {Ma}}{{Fakhouri} \&
  {Ma}}{2010}]{2010MNRAS.401.2245F}
{Fakhouri} O.,  {Ma} C.,  2010, \mnras, 401, 2245

\bibitem[\protect\citeauthoryear{{Fakhouri}, {Ma} \&
  {Boylan-Kolchin}}{{Fakhouri} et~al.}{2010}]{2010MNRAS.406.2267F}
{Fakhouri} O.,  {Ma} C.,    {Boylan-Kolchin} M.,  2010, \mnras, 406, 2267

\bibitem[\protect\citeauthoryear{{Fardal}, {Katz}, {Gardner}, {Hernquist},
  {Weinberg} \& {Dav{\'e}}}{{Fardal} et~al.}{2001}]{2001ApJ...562..605F}
{Fardal} M.~A.,  {Katz} N.,  {Gardner} J.~P.,  {Hernquist} L.,  {Weinberg}
  D.~H.,    {Dav{\'e}} R.,  2001, \apj, 562, 605

\bibitem[\protect\citeauthoryear{{Faucher-Gigu{\`e}re} \& {Kere{\v
  s}}}{{Faucher-Gigu{\`e}re} \& {Kere{\v s}}}{2011}]{2011MNRAS.tmpL.208F}
{Faucher-Gigu{\`e}re} C.-A.,  {Kere{\v s}} D.,  2011, \mnras, p.~L208

\bibitem[\protect\citeauthoryear{{Faucher-Gigu{\`e}re}, {Kere{\v s}},
  {Dijkstra}, {Hernquist} \& {Zaldarriaga}}{{Faucher-Gigu{\`e}re}
  et~al.}{2010}]{2010ApJ...725..633F}
{Faucher-Gigu{\`e}re} C.-A.,  {Kere{\v s}} D.,  {Dijkstra} M.,  {Hernquist} L.,
     {Zaldarriaga} M.,  2010, \apj, 725, 633

\bibitem[\protect\citeauthoryear{{Faucher-Gigu{\`e}re}, {Lidz}, {Hernquist} \&
  {Zaldarriaga}}{{Faucher-Gigu{\`e}re} et~al.}{2008a}]{2008ApJ...682L...9F}
{Faucher-Gigu{\`e}re} C.-A.,  {Lidz} A.,  {Hernquist} L.,    {Zaldarriaga} M.,
  2008a, \apjl, 682, L9

\bibitem[\protect\citeauthoryear{{Faucher-Gigu{\`e}re}, {Lidz}, {Hernquist} \&
  {Zaldarriaga}}{{Faucher-Gigu{\`e}re} et~al.}{2008b}]{2008ApJ...688...85F}
{Faucher-Gigu{\`e}re} C.-A.,  {Lidz} A.,  {Hernquist} L.,    {Zaldarriaga} M.,
  2008b, \apj, 688, 85

\bibitem[\protect\citeauthoryear{{Faucher-Gigu{\`e}re}, {Lidz}, {Zaldarriaga}
  \& {Hernquist}}{{Faucher-Gigu{\`e}re} et~al.}{2009}]{2009ApJ...703.1416F}
{Faucher-Gigu{\`e}re} C.-A.,  {Lidz} A.,  {Zaldarriaga} M.,    {Hernquist} L.,
  2009, \apj, 703, 1416

\bibitem[\protect\citeauthoryear{{Faucher-Gigu{\`e}re}, {Prochaska}, {Lidz},
  {Hernquist} \& {Zaldarriaga}}{{Faucher-Gigu{\`e}re}
  et~al.}{2008}]{2008ApJ...681..831F}
{Faucher-Gigu{\`e}re} C.-A.,  {Prochaska} J.~X.,  {Lidz} A.,  {Hernquist} L.,
   {Zaldarriaga} M.,  2008, \apj, 681, 831

\bibitem[\protect\citeauthoryear{{Forcada-Miro} \& {White}}{{Forcada-Miro} \&
  {White}}{1997}]{1997astro.ph.12204F}
{Forcada-Miro} M.~I.,  {White} S.~D.~M.,  1997, ArXiv Astrophysics e-prints

\bibitem[\protect\citeauthoryear{{F{\"o}rster Schreiber}, {Genzel},
  {Bouch{\'e}} \& {Cresci}}{{F{\"o}rster Schreiber}
  et~al.}{2009}]{2009ApJ...706.1364F}
{F{\"o}rster Schreiber} N.~M.,  {Genzel} R.,  {Bouch{\'e}} N.,    {Cresci} G.,
  2009, \apj, 706, 1364

\bibitem[\protect\citeauthoryear{{Genel}, {Bouch{\'e}}, {Naab}, {Sternberg} \&
  {Genzel}}{{Genel} et~al.}{2010}]{2010ApJ...719..229G}
{Genel} S.,  {Bouch{\'e}} N.,  {Naab} T.,  {Sternberg} A.,    {Genzel} R.,
  2010, \apj, 719, 229

\bibitem[\protect\citeauthoryear{{Genel}, {Genzel}, {Bouch{\'e}}, {Sternberg},
  {Naab}, {Schreiber}, {Shapiro}, {Tacconi}, {Lutz}, {Cresci}, {Buschkamp},
  {Davies} \& {Hicks}}{{Genel} et~al.}{2008}]{2008ApJ...688..789G}
{Genel} S.,  {Genzel} R.,  {Bouch{\'e}} N.,  {Sternberg} A.,  {Naab} T.,
  {Schreiber} N.~M.~F.,  {Shapiro} K.~L.,  {Tacconi} L.~J.,  {Lutz} D.,
  {Cresci} G.,  {Buschkamp} P.,  {Davies} R.~I.,    {Hicks} E.~K.~S.,  2008,
  \apj, 688, 789

\bibitem[\protect\citeauthoryear{{Genzel}, {Tacconi}, {Eisenhauer} \&
  {F{\"o}rster Schreiber}}{{Genzel} et~al.}{2006}]{2006Natur.442..786G}
{Genzel} R.,  {Tacconi} L.~J.,  {Eisenhauer} F.,    {F{\"o}rster Schreiber}
  N.~M.,  2006, \nat, 442, 786

\bibitem[\protect\citeauthoryear{{Genzel}, {Tacconi}, {Gracia-Carpio} \&
  {Sternberg}}{{Genzel} et~al.}{2010}]{2010MNRAS.407.2091G}
{Genzel} R.,  {Tacconi} L.~J.,  {Gracia-Carpio} J.,    {Sternberg} A.,  2010,
  \mnras, 407, 2091

\bibitem[\protect\citeauthoryear{{Genzel et al.}}{{Genzel et
  al.}}{2011}]{2011ApJ...733..101G}
{Genzel et al.} 2011, \apj, 733, 101

\bibitem[\protect\citeauthoryear{{Gingold} \& {Monaghan}}{{Gingold} \&
  {Monaghan}}{1977}]{1977MNRAS.181..375G}
{Gingold} R.~A.,  {Monaghan} J.~J.,  1977, \mnras, 181, 375

\bibitem[\protect\citeauthoryear{{Gnedin}}{{Gnedin}}{2000}]{2000ApJ...542..535G}
{Gnedin} N.~Y.,  2000, \apj, 542, 535

\bibitem[\protect\citeauthoryear{{Goerdt}, {Dekel}, {Sternberg}, {Ceverino},
  {Teyssier} \& {Primack}}{{Goerdt} et~al.}{2010}]{2010MNRAS.407..613G}
{Goerdt} T.,  {Dekel} A.,  {Sternberg} A.,  {Ceverino} D.,  {Teyssier} R.,
  {Primack} J.~R.,  2010, \mnras, 407, 613

\bibitem[\protect\citeauthoryear{{Hayward}, {Kere{\v s}}, {Jonsson},
  {Narayanan}, {Cox} \& {Hernquist}}{{Hayward}
  et~al.}{2011}]{2011arXiv1101.0002H}
{Hayward} C.~C.,  {Kere{\v s}} D.,  {Jonsson} P.,  {Narayanan} D.,  {Cox}
  T.~J.,    {Hernquist} L.,  2011, ArXiv e-prints

\bibitem[\protect\citeauthoryear{{Heckman}}{{Heckman}}{2002}]{2002ASPC..254..292H}
{Heckman} T.~M.,  2002, in {J.~S.~Mulchaey \& J.~T.~Stocke} ed., Extragalactic
  Gas at Low Redshift Vol.~254 of Astronomical Society of the Pacific
  Conference Series, {Galactic Superwinds Circa 2001}.
p.~292

\bibitem[\protect\citeauthoryear{{Heckman}, {Lehnert}, {Strickland} \&
  {Armus}}{{Heckman} et~al.}{2000}]{2000ApJS..129..493H}
{Heckman} T.~M.,  {Lehnert} M.~D.,  {Strickland} D.~K.,    {Armus} L.,  2000,
  \apjs, 129, 493

\bibitem[\protect\citeauthoryear{{Hernquist}}{{Hernquist}}{1987}]{1987ApJS...64..715H}
{Hernquist} L.,  1987, \apjs, 64, 715

\bibitem[\protect\citeauthoryear{{Hockney} \& {Eastwood}}{{Hockney} \&
  {Eastwood}}{1988}]{1988csup.book.....H}
{Hockney} R.~W.,  {Eastwood} J.~W.,  1988, {Computer simulation using
  particles}.
Bristol: Hilger, 1988

\bibitem[\protect\citeauthoryear{{Hoekstra}, {Hsieh}, {Yee}, {Lin} \&
  {Gladders}}{{Hoekstra} et~al.}{2005}]{2005ApJ...635...73H}
{Hoekstra} H.,  {Hsieh} B.~C.,  {Yee} H.~K.~C.,  {Lin} H.,    {Gladders} M.~D.,
   2005, \apj, 635, 73

\bibitem[\protect\citeauthoryear{{Hopkins}, {Cox}, {Kere{\v s}} \&
  {Hernquist}}{{Hopkins} et~al.}{2008}]{2008ApJS..175..390H}
{Hopkins} P.~F.,  {Cox} T.~J.,  {Kere{\v s}} D.,    {Hernquist} L.,  2008,
  \apjs, 175, 390

\bibitem[\protect\citeauthoryear{{Hopkins}, {Quataert} \& {Murray}}{{Hopkins}
  et~al.}{2011}]{2011arXiv1101.4940H}
{Hopkins} P.~F.,  {Quataert} E.,    {Murray} N.,  2011, ArXiv e-prints

\bibitem[\protect\citeauthoryear{{Hopkins et al.}}{{Hopkins et
  al.}}{2010}]{2010ApJ...724..915H}
{Hopkins et al.} 2010, \apj, 724, 915

\bibitem[\protect\citeauthoryear{{Katz}, {Keres}, {Dave} \& {Weinberg}}{{Katz}
  et~al.}{2003}]{2003ASSL..281..185K}
{Katz} N.,  {Keres} D.,  {Dave} R.,    {Weinberg} D.~H.,  2003, in
  {J.~L.~Rosenberg \& M.~E.~Putman} ed., The IGM/Galaxy Connection. The
  Distribution of Baryons at z=0 Vol.~281 of Astrophysics and Space Science
  Library, {How Do Galaxies Get Their Gas?}.
p.~185

\bibitem[\protect\citeauthoryear{{Katz}, {Weinberg} \& {Hernquist}}{{Katz}
  et~al.}{1996}]{1996ApJS..105...19K}
{Katz} N.,  {Weinberg} D.~H.,    {Hernquist} L.,  1996, \apjs, 105, 19

\bibitem[\protect\citeauthoryear{{Kennicutt}
  Jr.}{{Kennicutt}}{1998}]{1998ApJ...498..541K}
{Kennicutt} Jr. R.~C.,  1998, \apj, 498, 541

\bibitem[\protect\citeauthoryear{{Kere{\v s}} \& {Hernquist}}{{Kere{\v s}} \&
  {Hernquist}}{2009}]{2009ApJ...700L...1K}
{Kere{\v s}} D.,  {Hernquist} L.,  2009, \apjl, 700, L1

\bibitem[\protect\citeauthoryear{{Kere{\v s}}, {Katz}, {Weinberg} \&
  {Dav{\'e}}}{{Kere{\v s}} et~al.}{2005}]{2005MNRAS.363....2K}
{Kere{\v s}} D.,  {Katz} N.,  {Weinberg} D.~H.,    {Dav{\'e}} R.,  2005,
  \mnras, 363, 2

\bibitem[\protect\citeauthoryear{{Kere{\v s} et al.}}{{Kere{\v s} et
  al.}}{2009a}]{2009MNRAS.395..160K}
{Kere{\v s} et al.} 2009a, \mnras, 395, 160

\bibitem[\protect\citeauthoryear{{Kere{\v s} et al.}}{{Kere{\v s} et
  al.}}{2009b}]{2009MNRAS.396.2332K}
{Kere{\v s} et al.} 2009b, \mnras, 396, 2332

\bibitem[\protect\citeauthoryear{{Kere\v{s}}}{{Kere\v{s}}}{2007}]{keres_thesis}
{Kere\v{s}} D.,  2007, PhD thesis, Univ. Massachusetts, Amherst

\bibitem[\protect\citeauthoryear{{Kere\v{s} et al.}}{{Kere\v{s} et
  al.}}{prep}]{keres_arepo}
{Kere\v{s} et al.} {in prep.}

\bibitem[\protect\citeauthoryear{{Khochfar} \& {Silk}}{{Khochfar} \&
  {Silk}}{2009}]{2009ApJ...700L..21K}
{Khochfar} S.,  {Silk} J.,  2009, \apjl, 700, L21

\bibitem[\protect\citeauthoryear{{Kimm}, {Slyz}, {Devriendt} \&
  {Pichon}}{{Kimm} et~al.}{2011}]{2011MNRAS.413L..51K}
{Kimm} T.,  {Slyz} A.,  {Devriendt} J.,    {Pichon} C.,  2011, \mnras, 413, L51

\bibitem[\protect\citeauthoryear{{Komatsu}, {Dunkley}, {Nolta}, {Bennett},
  {Gold}, {Hinshaw}, {Jarosik}, {Larson}, {Limon}, {Page}, {Spergel},
  {Halpern}, {Hill}, {Kogut}, {Meyer}, {Tucker}, {Weiland}, {Wollack} \&
  {Wright}}{{Komatsu} et~al.}{2009}]{2009ApJS..180..330K}
{Komatsu} E.,  {Dunkley} J.,  {Nolta} M.~R.,  {Bennett} C.~L.,  {Gold} B.,
  {Hinshaw} G.,  {Jarosik} N.,  {Larson} D.,  {Limon} M.,  {Page} L.,
  {Spergel} D.~N.,  {Halpern} M.,  {Hill} R.~S.,  {Kogut} A.,  {Meyer} S.~S.,
  {Tucker} G.~S.,  {Weiland} J.~L.,  {Wollack} E.,    {Wright} E.~L.,  2009,
  \apjs, 180, 330

\bibitem[\protect\citeauthoryear{{Kriek et al.}}{{Kriek et
  al.}}{2006}]{2006ApJ...649L..71K}
{Kriek et al.} 2006, \apjl, 649, L71

\bibitem[\protect\citeauthoryear{{Krumholz} \& {Burkert}}{{Krumholz} \&
  {Burkert}}{2010}]{2010ApJ...724..895K}
{Krumholz} M.,  {Burkert} A.,  2010, \apj, 724, 895

\bibitem[\protect\citeauthoryear{{Law}, {Steidel}, {Erb}, {Larkin}, {Pettini},
  {Shapley} \& {Wright}}{{Law} et~al.}{2009}]{2009ApJ...697.2057L}
{Law} D.~R.,  {Steidel} C.~C.,  {Erb} D.~K.,  {Larkin} J.~E.,  {Pettini} M.,
  {Shapley} A.~E.,    {Wright} S.~A.,  2009, \apj, 697, 2057

\bibitem[\protect\citeauthoryear{{Leitner} \& {Kravtsov}}{{Leitner} \&
  {Kravtsov}}{2011}]{2011ApJ...734...48L}
{Leitner} S.~N.,  {Kravtsov} A.~V.,  2011, \apj, 734, 48

\bibitem[\protect\citeauthoryear{{Lu}, {Kere{\v s}}, {Katz}, {Mo}, {Fardal} \&
  {Weinberg}}{{Lu} et~al.}{2011}]{2011MNRAS.tmp.1035L}
{Lu} Y.,  {Kere{\v s}} D.,  {Katz} N.,  {Mo} H.~J.,  {Fardal} M.,    {Weinberg}
  M.~D.,  2011, \mnras, pp 1035--+

\bibitem[\protect\citeauthoryear{{Lucy}}{{Lucy}}{1977}]{1977AJ.....82.1013L}
{Lucy} L.~B.,  1977, \aj, 82, 1013

\bibitem[\protect\citeauthoryear{{Maller}, {Katz}, {Kere{\v s}}, {Dav{\'e}} \&
  {Weinberg}}{{Maller} et~al.}{2006}]{2006ApJ...647..763M}
{Maller} A.~H.,  {Katz} N.,  {Kere{\v s}} D.,  {Dav{\'e}} R.,    {Weinberg}
  D.~H.,  2006, \apj, 647, 763

\bibitem[\protect\citeauthoryear{{Martin} \& {Bouch{\'e}}}{{Martin} \&
  {Bouch{\'e}}}{2009}]{2009ApJ...703.1394M}
{Martin} C.~L.,  {Bouch{\'e}} N.,  2009, \apj, 703, 1394

\bibitem[\protect\citeauthoryear{{McBride}, {Fakhouri} \& {Ma}}{{McBride}
  et~al.}{2009}]{2009MNRAS.398.1858M}
{McBride} J.,  {Fakhouri} O.,    {Ma} C.,  2009, \mnras, 398, 1858

\bibitem[\protect\citeauthoryear{{McCarthy}, {Schaye}, {Ponman}, {Bower},
  {Booth}, {Dalla Vecchia}, {Crain}, {Springel}, {Theuns} \&
  {Wiersma}}{{McCarthy} et~al.}{2010}]{2010MNRAS.406..822M}
{McCarthy} I.~G.,  {Schaye} J.,  {Ponman} T.~J.,  {Bower} R.~G.,  {Booth}
  C.~M.,  {Dalla Vecchia} C.,  {Crain} R.~A.,  {Springel} V.,  {Theuns} T.,
  {Wiersma} R.~P.~C.,  2010, \mnras, 406, 822

\bibitem[\protect\citeauthoryear{{McGaugh}}{{McGaugh}}{2005}]{2005ApJ...632..859M}
{McGaugh} S.~S.,  2005, \apj, 632, 859

\bibitem[\protect\citeauthoryear{{Mihos} \& {Hernquist}}{{Mihos} \&
  {Hernquist}}{1996}]{1996ApJ...464..641M}
{Mihos} J.~C.,  {Hernquist} L.,  1996, \apj, 464, 641

\bibitem[\protect\citeauthoryear{{Murali}, {Katz}, {Hernquist}, {Weinberg} \&
  {Dav{\'e}}}{{Murali} et~al.}{2002}]{2002ApJ...571....1M}
{Murali} C.,  {Katz} N.,  {Hernquist} L.,  {Weinberg} D.~H.,    {Dav{\'e}} R.,
  2002, \apj, 571, 1

\bibitem[\protect\citeauthoryear{{Murray}, {M{\'e}nard} \& {Thompson}}{{Murray}
  et~al.}{2011}]{2011ApJ...735...66M}
{Murray} N.,  {M{\'e}nard} B.,    {Thompson} T.~A.,  2011, \apj, 735, 66

\bibitem[\protect\citeauthoryear{{Murray}, {Quataert} \& {Thompson}}{{Murray}
  et~al.}{2005}]{2005ApJ...618..569M}
{Murray} N.,  {Quataert} E.,    {Thompson} T.~A.,  2005, \apj, 618, 569

\bibitem[\protect\citeauthoryear{{Neistein} \& {Dekel}}{{Neistein} \&
  {Dekel}}{2008}]{2008MNRAS.383..615N}
{Neistein} E.,  {Dekel} A.,  2008, \mnras, 383, 615

\bibitem[\protect\citeauthoryear{{Nordon et al.}}{{Nordon et
  al.}}{2010}]{2010A&A...518L..24N}
{Nordon et al.} 2010, \aap, 518, L24

\bibitem[\protect\citeauthoryear{{Ocvirk}, {Pichon} \& {Teyssier}}{{Ocvirk}
  et~al.}{2008}]{2008MNRAS.390.1326O}
{Ocvirk} P.,  {Pichon} C.,    {Teyssier} R.,  2008, \mnras, 390, 1326

\bibitem[\protect\citeauthoryear{{Oppenheimer} \& {Dav{\'e}}}{{Oppenheimer} \&
  {Dav{\'e}}}{2008}]{2008MNRAS.387..577O}
{Oppenheimer} B.~D.,  {Dav{\'e}} R.,  2008, \mnras, 387, 577

\bibitem[\protect\citeauthoryear{{Oppenheimer et al.}}{{Oppenheimer et
  al.}}{2010}]{2010MNRAS.406.2325O}
{Oppenheimer et al.} 2010, \mnras, 406, 2325

\bibitem[\protect\citeauthoryear{{Pettini}, {Shapley}, {Steidel}, {Cuby},
  {Dickinson}, {Moorwood}, {Adelberger} \& {Giavalisco}}{{Pettini}
  et~al.}{2001}]{2001ApJ...554..981P}
{Pettini} M.,  {Shapley} A.~E.,  {Steidel} C.~C.,  {Cuby} J.-G.,  {Dickinson}
  M.,  {Moorwood} A.~F.~M.,  {Adelberger} K.~L.,    {Giavalisco} M.,  2001,
  \apj, 554, 981

\bibitem[\protect\citeauthoryear{{Press} \& {Schechter}}{{Press} \&
  {Schechter}}{1974}]{1974ApJ...187..425P}
{Press} W.~H.,  {Schechter} P.,  1974, \apj, 187, 425

\bibitem[\protect\citeauthoryear{{Prochaska} \& {Wolfe}}{{Prochaska} \&
  {Wolfe}}{2009}]{2009ApJ...696.1543P}
{Prochaska} J.~X.,  {Wolfe} A.~M.,  2009, \apj, 696, 1543

\bibitem[\protect\citeauthoryear{{Rees} \& {Ostriker}}{{Rees} \&
  {Ostriker}}{1977}]{1977MNRAS.179..541R}
{Rees} M.~J.,  {Ostriker} J.~P.,  1977, \mnras, 179, 541

\bibitem[\protect\citeauthoryear{{Sancisi}, {Fraternali}, {Oosterloo} \& {van
  der Hulst}}{{Sancisi} et~al.}{2008}]{2008A&ARv..15..189S}
{Sancisi} R.,  {Fraternali} F.,  {Oosterloo} T.,    {van der Hulst} T.,  2008,
  \aapr, 15, 189

\bibitem[\protect\citeauthoryear{{Schaye}, {Dalla Vecchia}, {Booth}, {Wiersma},
  {Theuns}, {Haas}, {Bertone}, {Duffy}, {McCarthy} \& {van de Voort}}{{Schaye}
  et~al.}{2010}]{2010MNRAS.402.1536S}
{Schaye} J.,  {Dalla Vecchia} C.,  {Booth} C.~M.,  {Wiersma} R.~P.~C.,
  {Theuns} T.,  {Haas} M.~R.,  {Bertone} S.,  {Duffy} A.~R.,  {McCarthy} I.~G.,
     {van de Voort} F.,  2010, \mnras, 402, 1536

\bibitem[\protect\citeauthoryear{{Shapley}, {Steidel}, {Pettini} \&
  {Adelberger}}{{Shapley} et~al.}{2003}]{2003ApJ...588...65S}
{Shapley} A.~E.,  {Steidel} C.~C.,  {Pettini} M.,    {Adelberger} K.~L.,  2003,
  \apj, 588, 65

\bibitem[\protect\citeauthoryear{{Sheth} \& {Tormen}}{{Sheth} \&
  {Tormen}}{2002}]{2002MNRAS.329...61S}
{Sheth} R.~K.,  {Tormen} G.,  2002, \mnras, 329, 61

\bibitem[\protect\citeauthoryear{{Silk}}{{Silk}}{1977}]{1977ApJ...211..638S}
{Silk} J.,  1977, \apj, 211, 638

\bibitem[\protect\citeauthoryear{{Simha}, {Weinberg}, {Dav{\'e}}, {Gnedin},
  {Katz} \& {Kere{\v s}}}{{Simha} et~al.}{2009}]{2009MNRAS.399..650S}
{Simha} V.,  {Weinberg} D.~H.,  {Dav{\'e}} R.,  {Gnedin} O.~Y.,  {Katz} N.,
  {Kere{\v s}} D.,  2009, \mnras, 399, 650

\bibitem[\protect\citeauthoryear{{Somerville}, {Hopkins}, {Cox}, {Robertson} \&
  {Hernquist}}{{Somerville} et~al.}{2008}]{2008MNRAS.391..481S}
{Somerville} R.~S.,  {Hopkins} P.~F.,  {Cox} T.~J.,  {Robertson} B.~E.,
  {Hernquist} L.,  2008, \mnras, 391, 481

\bibitem[\protect\citeauthoryear{{Springel}}{{Springel}}{2005}]{2005MNRAS.364.1105S}
{Springel} V.,  2005, \mnras, 364, 1105

\bibitem[\protect\citeauthoryear{{Springel}}{{Springel}}{2010}]{2010MNRAS.401..791S}
{Springel} V.,  2010, \mnras, 401, 791

\bibitem[\protect\citeauthoryear{{Springel} \& {Hernquist}}{{Springel} \&
  {Hernquist}}{2002}]{2002MNRAS.333..649S}
{Springel} V.,  {Hernquist} L.,  2002, \mnras, 333, 649

\bibitem[\protect\citeauthoryear{{Springel} \& {Hernquist}}{{Springel} \&
  {Hernquist}}{2003a}]{2003MNRAS.339..289S}
{Springel} V.,  {Hernquist} L.,  2003a, \mnras, 339, 289

\bibitem[\protect\citeauthoryear{{Springel} \& {Hernquist}}{{Springel} \&
  {Hernquist}}{2003b}]{2003MNRAS.339..312S}
{Springel} V.,  {Hernquist} L.,  2003b, \mnras, 339, 312

\bibitem[\protect\citeauthoryear{{Springel}, {White}, {Jenkins}, {Frenk},
  {Yoshida}, {Gao}, {Navarro}, {Thacker}, {Croton}, {Helly}, {Peacock}, {Cole},
  {Thomas}, {Couchman}, {Evrard}, {Colberg} \& {Pearce}}{{Springel}
  et~al.}{2005}]{2005Natur.435..629S}
{Springel} V.,  {White} S.~D.~M.,  {Jenkins} A.,  {Frenk} C.~S.,  {Yoshida} N.,
   {Gao} L.,  {Navarro} J.,  {Thacker} R.,  {Croton} D.,  {Helly} J.,
  {Peacock} J.~A.,  {Cole} S.,  {Thomas} P.,  {Couchman} H.,  {Evrard} A.,
  {Colberg} J.,    {Pearce} F.,  2005, \nat, 435, 629

\bibitem[\protect\citeauthoryear{{Steidel}, {Erb}, {Shapley}, {Pettini},
  {Reddy}, {Bogosavljevi{\'c}}, {Rudie} \& {Rakic}}{{Steidel}
  et~al.}{2010}]{2010ApJ...717..289S}
{Steidel} C.~C.,  {Erb} D.~K.,  {Shapley} A.~E.,  {Pettini} M.,  {Reddy} N.,
  {Bogosavljevi{\'c}} M.,  {Rudie} G.~C.,    {Rakic} O.,  2010, \apj, 717, 289

\bibitem[\protect\citeauthoryear{{Steidel}, {Shapley}, {Pettini}, {Adelberger},
  {Erb}, {Reddy} \& {Hunt}}{{Steidel} et~al.}{2004}]{2004ApJ...604..534S}
{Steidel} C.~C.,  {Shapley} A.~E.,  {Pettini} M.,  {Adelberger} K.~L.,  {Erb}
  D.~K.,  {Reddy} N.~A.,    {Hunt} M.~P.,  2004, \apj, 604, 534

\bibitem[\protect\citeauthoryear{{Stewart}, {Kaufmann}, {Bullock}, {Barton},
  {Maller}, {Diemand} \& {Wadsley}}{{Stewart}
  et~al.}{2011}]{2011ApJ...735L...1S}
{Stewart} K.~R.,  {Kaufmann} T.,  {Bullock} J.~S.,  {Barton} E.~J.,  {Maller}
  A.~H.,  {Diemand} J.,    {Wadsley} J.,  2011, \apjl, 735, L1+

\bibitem[\protect\citeauthoryear{{Thoul} \& {Weinberg}}{{Thoul} \&
  {Weinberg}}{1996}]{1996ApJ...465..608T}
{Thoul} A.~A.,  {Weinberg} D.~H.,  1996, \apj, 465, 608

\bibitem[\protect\citeauthoryear{{van de Voort}, {Schaye}, {Booth}, {Haas} \&
  {Dalla Vecchia}}{{van de Voort} et~al.}{2011}]{2011MNRAS.414.2458V}
{van de Voort} F.,  {Schaye} J.,  {Booth} C.~M.,  {Haas} M.~R.,    {Dalla
  Vecchia} C.,  2011, \mnras, 414, 2458

\bibitem[\protect\citeauthoryear{{van Dokkum et al.}}{{van Dokkum et
  al.}}{2006}]{2006ApJ...638L..59V}
{van Dokkum et al.} 2006, \apjl, 638, L59

\bibitem[\protect\citeauthoryear{{Wechsler}, {Bullock}, {Primack}, {Kravtsov}
  \& {Dekel}}{{Wechsler} et~al.}{2002}]{2002ApJ...568...52W}
{Wechsler} R.~H.,  {Bullock} J.~S.,  {Primack} J.~R.,  {Kravtsov} A.~V.,
  {Dekel} A.,  2002, \apj, 568, 52

\bibitem[\protect\citeauthoryear{{White}}{{White}}{2002}]{2002ApJS..143..241W}
{White} M.,  2002, \apjs, 143, 241

\bibitem[\protect\citeauthoryear{{White} \& {Frenk}}{{White} \&
  {Frenk}}{1991}]{1991ApJ...379...52W}
{White} S.~D.~M.,  {Frenk} C.~S.,  1991, \apj, 379, 52

\bibitem[\protect\citeauthoryear{{White} \& {Rees}}{{White} \&
  {Rees}}{1978}]{1978MNRAS.183..341W}
{White} S.~D.~M.,  {Rees} M.~J.,  1978, \mnras, 183, 341

\bibitem[\protect\citeauthoryear{{Wright}, {Larkin}, {Law}, {Steidel},
  {Shapley} \& {Erb}}{{Wright} et~al.}{2009}]{2009ApJ...699..421W}
{Wright} S.~A.,  {Larkin} J.~E.,  {Law} D.~R.,  {Steidel} C.~C.,  {Shapley}
  A.~E.,    {Erb} D.~K.,  2009, \apj, 699, 421

\end{thebibliography}
 
\end{document}